\documentclass[12pt]{article}
\hoffset= - 2cm
\usepackage{amssymb,graphicx}
\usepackage{epstopdf}
\usepackage{amsmath,amsfonts}
\usepackage[normalem]{ulem}
\usepackage{epsfig}
\usepackage[dvipsnames]{xcolor}

\def\eps{\epsilon}

\newcommand{\be}{\begin{equation}}
\newcommand{\ee}{\end{equation}}
\newcommand{\bea}{\begin{eqnarray}}
\newcommand{\eea}{\end{eqnarray}}
\newcommand{\bg}{\begin{gather}}
\newcommand{\eg}{\end{gather}}
\newcommand{\bseq}{\begin{subequations}}
\newcommand{\eseq}{\end{subequations}}

\usepackage{hyperref}
\definecolor{linkcolor}{HTML}{799B03}
\definecolor{urlcolor}{HTML}{799B03}
\hypersetup{pdfstartview=FitH,linkcolor=linkcolor,urlcolor=urlcolor,colorlinks=true}

\makeatletter
\newcommand*{\myfnsymbolsingle}[1]{%
  \ensuremath{%
    \ifcase#1
    \or 
      *%
    \or 
      \dagger
    \or 
      \ddagger
    \or 
      1
    \or 
      2
    \or
      3
    \or
      4
    \or
      5
    \or
      6
    \or
      7
    \or
      8
    \else 
      \@ctrerr
    \fi
  }%
}
\makeatother

\usepackage{alphalph}
\newalphalph{\myfnsymbolmult}[mult]{\myfnsymbolsingle}{}
\renewcommand*{\thefootnote}{%
  \myfnsymbolmult{\value{footnote}}%
}

\begin{document}

\begin{center}
  {\LARGE \bf  Subluminal cosmological bounce beyond Horndeski}

\vspace{10pt}

\vspace{20pt}
S. Mironov$^{a,c,d,e}$\footnote{sa.mironov\_1@physics.msu.ru},
V. Rubakov$^{a,b}$\footnote{rubakov@inr.ac.ru},
V. Volkova$^{a}$\footnote{volkova.viktoriya@physics.msu.ru}
\renewcommand*{\thefootnote}{\arabic{footnote}}
\vspace{15pt}

$^a$\textit{Institute for Nuclear Research of the Russian Academy of Sciences,\\
60th October Anniversary Prospect, 7a, 117312 Moscow, Russia}\\
\vspace{5pt}

$^b$\textit{Department of Particle Physics and Cosmology, Physics Faculty,\\
M.V. Lomonosov Moscow State University,\\
Vorobjevy Gory, 119991 Moscow, Russia}

$^c$\textit{Institute for Theoretical and Experimental Physics,\\
Bolshaya Cheriomyshkinskaya, 25, 117218 Moscow, Russia}

$^d$\textit{Moscow Institute of Physics and Technology,\\
Institutski pereulok, 9, 141701, Dolgoprudny, Russia}

$^e$\textit{Institute for Theoretical and Mathematical Physics,\\
M.V. Lomonosov Moscow State University, 119991 Moscow, Russia}
\end{center}

\vspace{5pt}

\begin{abstract} 
We address the issue of 
potential superluminal propagation of gravitational waves
in backgrounds
{neighboring} the previously suggested 
bounce~\cite{bounceI} in beyond Horndeski theory.
We find that the bouncing solution lies right at the
boundary of the region where the gravitational waves 
propagate at speed exceeding that of light,
i.e. that solution suffers superluminality problem.
We suggest a novel version of a completely stable bouncing model
where both scalar and tensor 
perturbations remain safely subluminal not only on the solution 
itself but also in its neighbourhood.
{The model remains free of superluminality
  when extra matter in the form of radiation or, more generally,
  ideal fluid
  with equation of state parameter $w\leq 1/3$ (and also somewhat
  higher)
  is added. Superluminality reappears when extra matter
  is added whose  sound velocity is equal or close to 1 in flat space;
  an example
  is scalar field minimally coupled to metric. The latter
  property is characteristic of all beyond Horndeski cosmologies; we briefly
discuss its significance.}

\end{abstract}

\section{Introduction}
\label{sec:intro}

Recent studies have shown that
Horndeski theories~\cite{Horndeski} and their
extensions~\cite{Zuma,Gleyzes}
offer a remarkable framework for
tackling various cosmological issues such as
late time accelerated expansion of the Universe,
or, notably, initial singularity problem
(for a review see, e.g., Ref.~\cite{KobaRev}).
A particularly attractive feature of these scalar-tensor 
theories is
their ability to allow for the Null Energy Condition (NEC) violation
\footnote{In fact, when gravity is modified, the NEC is replaced with
the Null Convergence Condition (NCC)~\cite{NCC}.}
with no catastrophic consequences for the stability of the solutions,
as reviewed, e.g., in
Ref.~\cite{RubakovNEC}.
For this reason, Horndeski and beyond Horndeski
theories have been extensively used 
for constructing spatially flat FLRW
cosmological scenarios with non-standard
dynamics, e.g., the bouncing Universe and the Universe starting off
with Genesis (see Refs.~\cite{KobaRev,Khalat} and references therein).

However, 
bouncing and Genesis models in unextended Horndeski 
theories still have severe problems related to stability. 
Namely, in these theories, the absence of
ghosts and gradient
instabilities {\itshape at all times} in a bouncing or Genesis
cosmology
can be achieved at 
a price of potential 
strong coupling and/or geodesic incompleteness; examples are given, 
e.g., in Refs.~\cite{LMR,Koba_nogo,IjjasBounce} and the no-go theorem is
proven
in
Refs.~\cite{LMR,Koba_nogo}.
What saved the day was beyond Horndeski theories where
geodesically complete bouncing and Genesis solutions were
constructed;
these solutions are stable during entire 
evolution without 
the risk of strong coupling~\cite{Cai,CreminelliBH,RomaBounce,CaiBounce,bounceI,genesisGR,chineseBounce1}. 
Similar construction~\cite{chineseBounce1,chineseBounce2} was given in
the context of more general DHOST theories~\cite{DHOST1,DHOST2}.

Another issue to worry about in modified gravities, including
Horndeski theories and their extensions, is the danger of superluminality.
This issue has been discussed from various viewpoints in
Refs.~\cite{superlum1,BabVikMukh,gen_original,subl_gen,MatMat,Unbraiding} 
and references therein. It is generally accepted that superluminal
propagation in any background is an undesirable feature 
that signals that the theory cannot descend, as low energy
effective field theory,
from any UV-complete, Lorentz-covariant theory
\footnote{In this paper, as well as in most other papers where
the (super)luminality issue is discussed, it is assumed that ordinary
light propagates in metric $g_{\mu\nu}$ 
(entering the action~\eqref{eq:lagrangian}). This point
is not entirely trivial, because the light cones are different for
metric $g_{\mu\nu}$ and for $\tilde{g}_{\mu\nu}$ related to
$g_{\mu\nu}$  by disformal transformation~\cite{Bekenstein}
$\tilde{g}_{\mu\nu} = g_{\mu\nu} + \Gamma(\pi,X)\partial_{\mu}\pi\partial_{\nu}\pi$. Therefore, the notion of Lorentz covariance 
needs qualification: it refers here to the theory where light feels
the metric $g_{\mu\nu}$.
}.
In practice (and in this paper in particular),
one often does not pretend to define the Lagrangian in the entire
phase space; one usually  keeps only those terms in
the Lagrangian that are sufficient for obtaining
the solution and analyzing its stability (i.e., terms that do not
vanish on the pertinent solution
and its
close
{neighborhood}).
In that case, {the minimal requirement is
the absence of superluminality} 
for perturbations about the cosmological solution of interest
{and
  in its vicinity. If this requirement is satisfied, one may become more
  ambitious and ask whether or not superluminality 
appears
  upon adding extra matter (say, ideal fluid), still within the
  domain of the phase space where 
  the 
  Lagrangian is known. This is a non-trivial issue:
  at least in the subluminal Genesis model
  of Ref.~\cite{subl_gen}, perturbations do become superluminal 
  in the presence of additional matter~\cite{MatMat}.
} 

It is worth noting that the superluminality and  stability problems are
not directly
related  to each other: superluminality may occur in a vicinity of
a stable part of solution (like in Refs.~\cite{gen_original,Unbraiding,IjjasBounceVik}).
So, the superluminality issue requires separate analysis in any
cosmological model in (beyond) Horndeski theory.

Recent examples of stable bouncing and Genesis solutions given in
Refs.~\cite{bounceI,genesisGR}
have been constructed in such
a way that there are no superluminal modes about the
solution during entire evolution.
However, the superluminality issue has not been analyzed even
in the vicinity of these solutions.
{One of the purposes of
this paper is to} fill this gap.
Namely,
we calculate the speed of perturbations about the homogeneous
and isotropic background in the phase space
around the completely stable bouncing solution
of Ref.~\cite{bounceI} and show that in the vicinity of this solution,
there exists  a
domain of the phase space
where tensor modes are superluminal. So, the bouncing solution
of Ref.~\cite{bounceI} does suffer superluminality problem.

This
{drawback can be cured}. 
We modify the Lagrangian  of Ref.~\cite{bounceI} in such a way that 
the bouncing solution still exists and is stable throughout the entire
evolution, while
both scalar and tensor sectors are safely subluminal
at all times in
a vicinity of 
the solution.
{Like the original
 {scenario}~\cite{bounceI},
  our new model tends to General Relativity with conventional massless
  scalar field in both early and late time asymptotics.}
It is worth noting that there are
examples of beyond Horndeski 
and DHOST models with stable bounces where the speed of tensor modes 
is identically equal to 1~\cite{chineseBounce1,chineseBounce2}.

{Our {new} 
model turns out to be 
  subluminal in  a broader context outlined above.
  Namely, when matter in the form of ideal fluid is added, there are
  no
  superluminal perturbations about the backgrounds for which we
  trust our Lagrangian, provided the equation of state parameter of
  the fluid is $w\leq 1/3$ and even somewhat larger.
  In this respect our model outperforms
  the subluminal Genesis model with cubic Galileon~\cite{subl_gen}.
  However, if the extra matter is added whose flat-space
  sound speed is equal or close
  to 1, superluminality reappears. In fact, the latter property holds for 
  all
  beyond Horndeski theories in {a} cosmological setting, no matter what sort
  of cosmological solution one considers. In particular, when
  one adds conventional scalar field minimally coupled to metric, and makes
  its energy density non-zero, perturbations of this scalar
  field mix with beyond Horndeski sector in such a way that their
  sound speed exceeds that of light. Presumably, this means that
  when a scalar-tensor gravity is in beyond Horndeski regime,
  all additional scalar fields also feature non-trivial properties such as
  unconventional kinetic
  terms in the Lagrangian, non-minimal coupling to gravity,
  etc.}

This paper is organized as follows. In Sec.~\ref{sec:setup}
we introduce the Lagrangian of beyond Horndeski theory considered in
Ref.~\cite{bounceI} and revisit the
stability and subluminality conditions. 
In Sec.~\ref{sec:superluminal} we study the bouncing solution 
of Ref.~\cite{bounceI} and
show that the gravitational waves
exhibit superluminal propagation in the vicinity of the solution.
In Sec.~\ref{sec:solution} we construct a modified 
version of a completely stable bouncing model
where
both scalar and tensor perturbations remain safely subluminal
not only on the solution itself but also in its neighbourhood.
{We add matter (ideal fluid)
to beyond Horndeski theory in Sec.~\ref{sec:extra_matter}
and derive both stability conditions and expressions for the sound speeds;
we then study the model of Sec.~\ref{sec:solution} and find that there are no
superluminality in the regions of the phase space where we trust our
Lagrangian, provided the equation of state parameter of
the fluid is $w\leq 1/3$ and even somewhat larger. 
We also point out that
in any beyond Horndeski cosmology,
one of the {scalar} modes is superluminal when one adds matter whose
flat-space sound speed
is close or equal to 1.}
We conclude in Sec.~\ref{sec:conclusion}.

\section{Superluminality near the original bouncing solution}

\subsection{Stability and subluminality conditions in beyond Horndeski theory}
\label{sec:setup}

Here we set up the notations and recall the general form of the
stability conditions for cosmological solutions
in (beyond) Horndeski theory~\cite{KobayashiG,RomaBounce,bounceI}.
Like in Ref.~\cite{bounceI}, it is sufficient for our purposes
to consider a subclass of beyond Horndeski theory,
whose Lagrangian reads
\begin{multline}
\label{eq:lagrangian}
\mathcal{L}(F,G_4,F_4) = F(\pi,X)
- G_4(\pi,X)R \\
+ \left(2 G_{4X}(\pi,X) - F_4 (\pi,X) \; X\right)\left[\left(\Box\pi\right)^2-\pi_{;\mu\nu}\pi^{;\mu\nu}\right] \\
+ 2 F_4 (\pi,X) \left[\pi^{,\mu} \pi_{;\mu\nu} \pi^{,\nu}\Box\pi -  \pi^{,\mu} \pi_{;\mu\lambda} \pi^{;\nu\lambda}\pi_{,\nu} \right],
\end{multline}
where $\pi$ is the scalar field,
$X=g^{\mu\nu}\pi_{,\mu}\pi_{,\nu}$,
$\pi_{,\mu}=\partial_\mu\pi$,
$\pi_{;\mu\nu}=\triangledown_\nu\triangledown_\mu\pi$,
$\Box\pi = g^{\mu\nu}\triangledown_\nu\triangledown_\mu\pi$,
$G_{4X}=\partial G_4/\partial X$. The functions $F$ and $G_4$
are characteristic of the  Horndeski theories, while
non-vanishing $F_4$  extends the theory to beyond Horndeski
type. In a cosmological context,  we consider  a homogeneous
background scalar field $\pi=\pi(t)$
and a spatially flat FLRW background metric 
\begin{equation}
\label{eq:FLRW}
\mathrm{d}s^2 = \mathrm{d}t^2 - a^2(t)\delta_{ij}\mathrm{d}x^i\mathrm{d}x^j.
\end{equation}
The gravitational equations for the
Lagrangian~\eqref{eq:lagrangian} read:
\begin{subequations}
\label{eq:Einstein}
\begin{align}
\label{eq:dg00}
\delta g^{00}: \;\;
&F-2F_XX+6H^2G_4+6HG_{4\pi}\dot{\pi}
-24H^2X(G_{4X}+G_{4XX}X)
+12HG_{4\pi X}X\dot{\pi}
\\\nonumber&
+6H^2X^2(5F_4+2F_{4X}X) = 0,\\
\label{eq:dgii}
\delta g^{ii}: \;\;
&F+2(3H^2+2\dot{H})G_4-12H^2G_{4X}X
-8\dot{H}G_{4X}X-8HG_{4X}\ddot{\pi}\dot{\pi}-16HG_{4XX}X\ddot{\pi}\dot{\pi}\quad
\\\nonumber&+2(\ddot{\pi}+2H\dot{\pi})G_{4\pi}+4XG_{4\pi X}(\ddot{\pi}-2H\dot{\pi})+2XG_{4\pi\pi}
 +2F_4X(3H^2X+2\dot{H}X+8H\ddot{\pi}\dot{\pi})\quad
 \\ \nonumber&+8HF_{4X}X^2\ddot{\pi}\dot{\pi}
+4HF_{4\pi}X^2\dot{\pi} = 0,
\end{align}
\end{subequations}
where $H=\dot{a}/a$ is the Hubble parameter. The scalar field equation
is a consequence
of eqs.~\eqref{eq:Einstein}.

To find whether perturbations about the homogeneous and isotropic
solution are stable and subluminal, one considers them at
a linearized level.
One adopts the ADM parametrization for the metric perturbations
\begin{equation}
\label{eq:FLRW_perturbed}
\mathrm{d}s^2 = N^2 \mathrm{d}t^2 - \gamma_{ij}(\mathrm{d}x^i+ N^i \mathrm{d}t)(\mathrm{d}x^j+N^j \mathrm{d}t),
\end{equation}
where 
\be
\label{eq:ADM}
N = 1+\alpha, \qquad N_i = \partial_i\beta, \qquad
\gamma_{ij}= a^2(t) e^{2\zeta} \left(\delta_{ij} + h_{ij}^T +
\dfrac12 h_{ik}^T {h^{k\:T}_j}\right).
\ee
In eq.~\eqref{eq:ADM} $\alpha$, $\beta$ and $\zeta$ belong to the 
scalar sector of perturbations, while $h_{ik}^T$ denote transverse
traceless tensor modes ($h_{ii}^T = 0, \partial_i h_{ij}^T = 0$). 
Below we
utilize the unitary gauge approach ($\delta\pi = 0$), where a dynamical
DOF in the scalar sector is the curvature perturbation $\zeta$.
Upon integrating out non-dynamical $\alpha$ and $\beta$, 
one arrives at
the unconstrained quadratic action 
for metric perturbations
(see Refs.~\cite{KobayashiG,RomaBounce,bounceI} for details)
\begin{equation}
\label{eq:quadratic_action}
S=\int\mathrm{d}t\mathrm{d}^3x \;a^3
\left[\dfrac{\mathcal{{G}_T}}{8}\left(\dot{h}^T_{ij}\right)^2
-\dfrac{\mathcal{F_T}}{8a^2}\left(\triangledown h_{ij}^T\right)^2
+\mathcal{G_S}\dot{\zeta}^2
-\mathcal{F_S}\dfrac{(\triangledown\zeta)^2}{a^2}\right],\\
\end{equation}
with
$(\triangledown\zeta)^2 = \delta^{ij} \partial_i \zeta \partial_j \zeta$,
$\triangle = \delta^{ij} \partial_i \partial_j$
and
\bea
\label{eq:GS_setup}
&&\mathcal{G_S}=\dfrac{\Sigma\mathcal{{G}_T}^2}{\Theta^2}+3\mathcal{{G}_T},\\
\label{eq:FS_setup}
&&\mathcal{F_S}=\dfrac{1}{a}\dfrac{\mathrm{d}\xi}{\mathrm{d}t}-\mathcal{F_T},\\
\label{eq:xi_func_setup}
&&\xi
=\dfrac{a\left(\mathcal{{G}_T}+\mathcal{D}\dot{\pi}\right)\mathcal{{G}_T}}{\Theta},
\eea
while the 
coefficients $\mathcal{G_T}$, $\mathcal{F_T}$, $\mathcal{D}$,
$\Theta$ and $\Sigma$ are expressed in terms of the Lagrangian functions
as follows:
\bea
\label{eq:GT_coeff_setup}
&&\mathcal{G_T}=2G_4-4G_{4X}X+ 2F_4X^2,
\qquad\qquad\mathcal{F_T}=2G_4,
\qquad\qquad\mathcal{D}=-2F_4X\dot{\pi},
\\
\label{eq:Theta_coeff_setup}
&&\Theta=
2G_4H-8HG_{4X}X-8HG_{4XX}X^2+G_{4\pi}\dot{\pi}+2G_{4\pi X}X\dot{\pi}
+10HF_4X^2+4HF_{4X}X^3
,\qquad\\
\label{eq:Sigma_coeff_setup}
&&\Sigma=F_XX+2F_{XX}X^2
-6H^2G_4+42H^2G_{4X}X+96H^2G_{4XX}X^2+24H^2G_{4XXX}X^3
\\
\nonumber&&
-6HG_{4\pi}\dot{\pi}  -30HG_{4\pi X}X\dot{\pi}-12HG_{4\pi XX}X^2\dot{\pi}
-90H^2F_4X^2-78H^2F_{4X}X^3
\\
\nonumber&&-12H^2F_{4XX}X^4.
\eea
The stability conditions for a cosmological solution
immediately follow from the expression for
the action~\eqref{eq:quadratic_action}:
\begin{equation}
\label{eq:stability_cond}
\mathcal{{G}_T}, \mathcal{F_T}> \eps >0,\quad \mathcal{G_S},
\mathcal{F_S} >\eps> 0 \;,
\end{equation}
where $\epsilon$ is a positive constant which ensures that
there is no naive strong coupling,
i.e.
$\mathcal{{G}_{S,T}} \not\to 0$
and/or $\mathcal{{F}_{S,T}} \not\to 0$
at any time including asymptotics $t\to \pm \infty$
(see Refs.~\cite{RomaBounce,bounceI} for discussion).
The absence of superluminal propagation at any time
is expressed in terms of sound speeds of scalar and tensor modes
as follows:
\be
\label{eq:sound_speeds}
c_\mathcal{T}^2=\dfrac{\mathcal{F_T}}{\mathcal{{G}_T}} \leq 1,\qquad
c_\mathcal{S}^2=\dfrac{\mathcal{F_S}}{\mathcal{G_S}} \leq 1.
\ee
As we discussed in Sec.~\ref{sec:intro}, the minimal subluminality
requirement is that these inequalities are satisfied for a
background solution of interest and nearby solutions to 
eqs.~\eqref{eq:Einstein} as well.

\subsection{Original bouncing solution and superluminality}
\label{sec:superluminal}
The bouncing model of Ref.~\cite{bounceI} complies with the
stability conditions~\eqref{eq:stability_cond} at all times,
thus, we claimed that it is {\itshape completely} stable. Also, the inequalities
\eqref{eq:sound_speeds} are satisfied at all times
      {\it for the bouncing solution itself}.
On top of that, the solution
has simple asymptotics as $t\to\pm\infty$: the beyond Horndeski theory
boils down to GR + conventional massless scalar field long before and
long after the bouncing epoch.
The latter property is possible only provided $\Theta$
in eq.~\eqref{eq:xi_func_setup} crosses zero at
some moment of time, see Ref.~\cite{bounceI}
for a detailed discussion. This is the so-called $\gamma$-crossing phenomenon
\footnote{The name of the phenomenon originates from
  Refs.~\cite{Ijjas,Unbraiding}, where the coefficient
  $\Theta$ is denoted by $\gamma$.},
which is completely harmless~\cite{BKLO,Ijjas}
and corresponds to
a transition between the
branches of solutions for the Hubble parameter $H$ in
eq.~\eqref{eq:dg00}~\cite{Unbraiding}.

However, a dangerous feature of the
bouncing solution of Ref.~\cite{bounceI}
is that
gravitational waves
propagate strictly at the
speed of light
throughout the entire evolution\footnote{Let
  us note that a setup  suggested within the EFT approach 
  in Ref.~\cite{CreminelliBH} may be also problematic due to
  strictly luminal propagation 
  of gravitational waves.}, i.e., $c_\mathcal{T}^2(t)=1$ for
all $t$.
The latter feature is due to a deliberate choice of
$\mathcal{{G}_T} =\mathcal{{F}_T} = 1$ for
all $ t$ on the solution,
which simplified the construction procedure.
This property may or may not signal that the
tensor perturbations are superluminal
in a vicinity of the
bouncing solution,
which is the case if the solution happens to be just
at the boundary (in phase space) between
regions with subluminal and superluminal gravitational waves.
Let us study this issue.

We begin with a brief description of 
the
solution of Ref.~\cite{bounceI}. It
has the following  Hubble parameter, with the bounce at $t=0$,
\be
\label{eq:Hubble}
H(t) = \frac{t}{3(\tau^2+t^2)}, \qquad a(t) = (\tau^2+t^2)^{\frac16},
\ee
where the parameter $\tau$ regulates the duration of the bouncing epoch.
The background scalar field was chosen as $\pi(t) = t$, which
is always possible to achieve {\it on a single solution } by
field redefinition. Hence,
in the FLRW background~\eqref{eq:FLRW} we have 
$X = g^{\mu\nu}\pi_{,\mu}\pi_{,\nu} = 1$ on this solution.
The explicit
bouncing solution of Ref.~\cite{bounceI} has been obtained by making use
of 
the so-called reconstruction procedure: for a chosen evolution
of the Hubble parameter~\eqref{eq:Hubble}, we
found an explicit Lagrangian of the form~\eqref{eq:lagrangian}
by making use of the stability conditions~\eqref{eq:stability_cond} 
and background gravitational equations~\eqref{eq:Einstein}
(see Ref.~\cite{bounceI} for a detailed description). 
Namely,
we take the Ansatz for the Lagrangian functions
in eq.~\eqref{eq:quadratic_action} in terms of powers of $X$:
\begin{subequations}
\label{eq:lagr_series}
\begin{align}
\label{F}
& F(\pi, X) = f_0(\pi) + f_1(\pi)\cdot X + f_2(\pi)\cdot X^2 \\
\label{G4}
& G_4(\pi, X) = \frac12 + g_{40}(\pi) + g_{41}(\pi) \cdot X,\\
\label{F4}
& F_4(\pi, X) = f_{40}(\pi) + f_{41}(\pi) \cdot X.
\end{align}
\end{subequations}
Note that this Ansatz defines the Lagrangian only on the solution
$X=1$ and its
close vicinity. 
The following steps amount to (i) choosing the functions $f_{4i}(\pi)$,
$g_{4i}(\pi)$ ($i=0,1$) and $f_2(\pi)$ in such a way  that
$\mathcal{{G}_T} =\mathcal{{F}_T} = 1$ and 
the stability
conditions~\eqref{eq:stability_cond} hold;
 (ii) the remaining functions $f_0(\pi)$ and $f_1(\pi)$ are found from
the
two independent background gravitational equations~\eqref{eq:Einstein}.
An additional requirement imposed on the Lagrangian functions in
Ref.~\cite{bounceI} was
that the beyond Horndeski
theory~\eqref{eq:lagrangian} reduces to
GR + conventional massless scalar field in both
asymptotic past and asymptotic future.
This implies that
$F(\pi, X) \to {X}/{(3\pi^2)}$ (the field $\phi\propto \log \pi$ is
a canonical massless scalar field),
$G_4(\pi, X) \to 1/2$ and
$F_4(\pi, X) \to 0$
as $\pi=t\to\pm\infty$  (hereafter we set $M_{Pl}^2/8\pi =1$).
Clearly, there remains substantial arbitrariness in the choice of the functions
$f_0 (\pi), \dots , f_{41} (\pi)$; one example is given in Appendix A.

\begin{figure}[h!]\begin{center}\hspace{-1cm}
\put(0,290){$\dot{\pi}$}
{\includegraphics[width=0.6\linewidth]{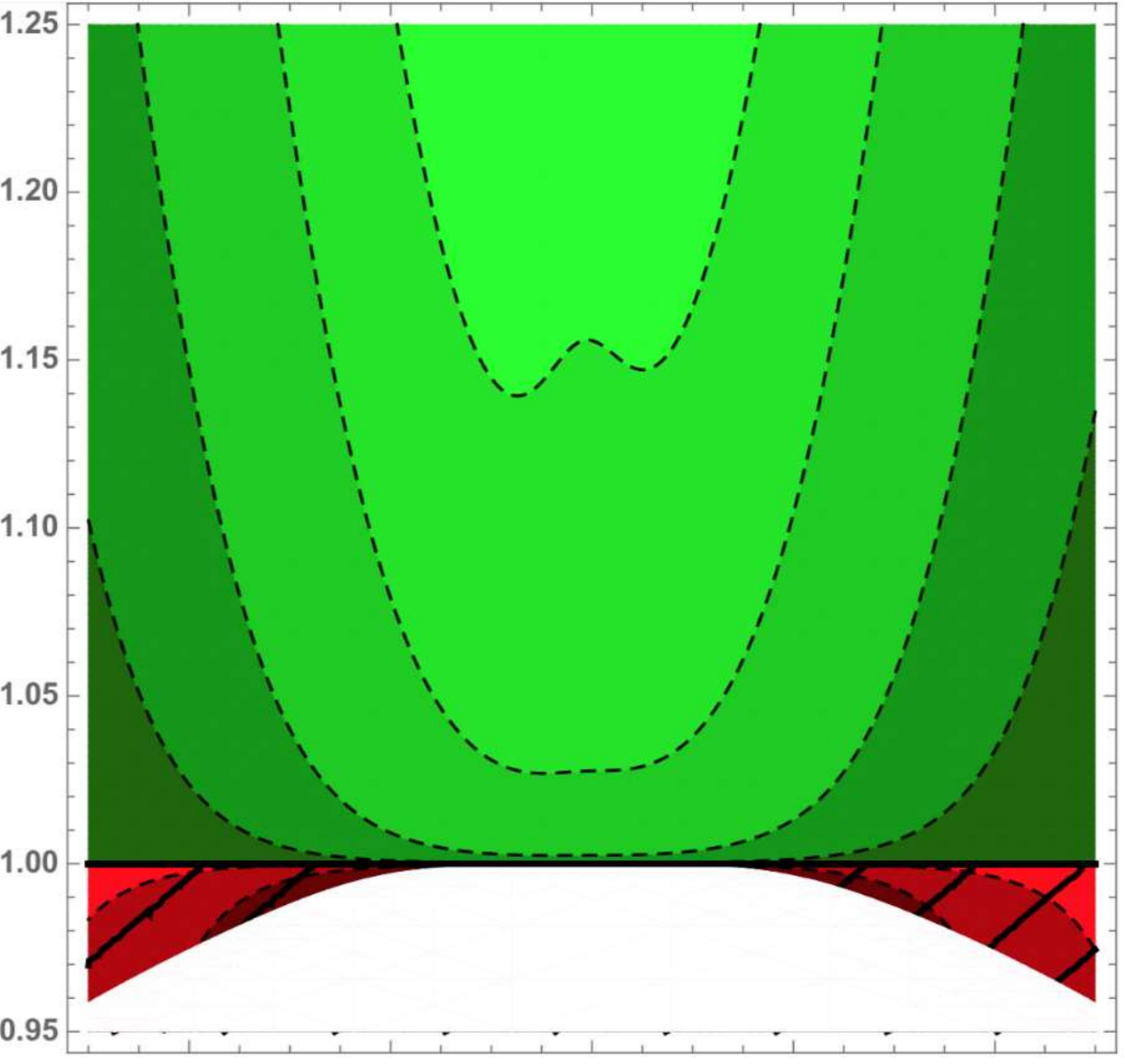}}
\hspace{0.4cm}
{\includegraphics[width=0.12\linewidth]{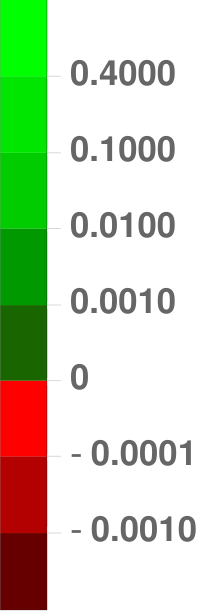}\hspace{-2.5cm}$\pi$}
\caption{[color online] Variance
  $(1 - c_\mathcal{T}^2(\pi,\dot{\pi}))$ of the speed squared of tensor
  modes in the
  phase space $(\pi,\dot{\pi})$ for the original model with 
bouncing solution~\cite{bounceI}. Dashed lines are lines of constant 
$(1 - c_\mathcal{T}^2(\pi,\dot{\pi}))$.
White region is the one where
the solution does not exist, see text.
A black horizontal line corresponds to the original bounce 
with $\pi(t)=t$ (hence, $\dot{\pi}=1$) 
and $c_\mathcal{T}^2 = 1$. Negative values of
 $(1 - c_\mathcal{T}^2(\pi,\dot{\pi}))$ (hatched region) mean superluminal propagation.
The original solution lies right on the verge of
the domain with superluminal tensor modes.
} \label{fig:tensor_old1}
\end{center}\end{figure}

\begin{figure}[h!]\begin{center}\hspace{-1cm}
\put(0,300){$\dot{\pi}$}
{\includegraphics[width=0.6\linewidth]{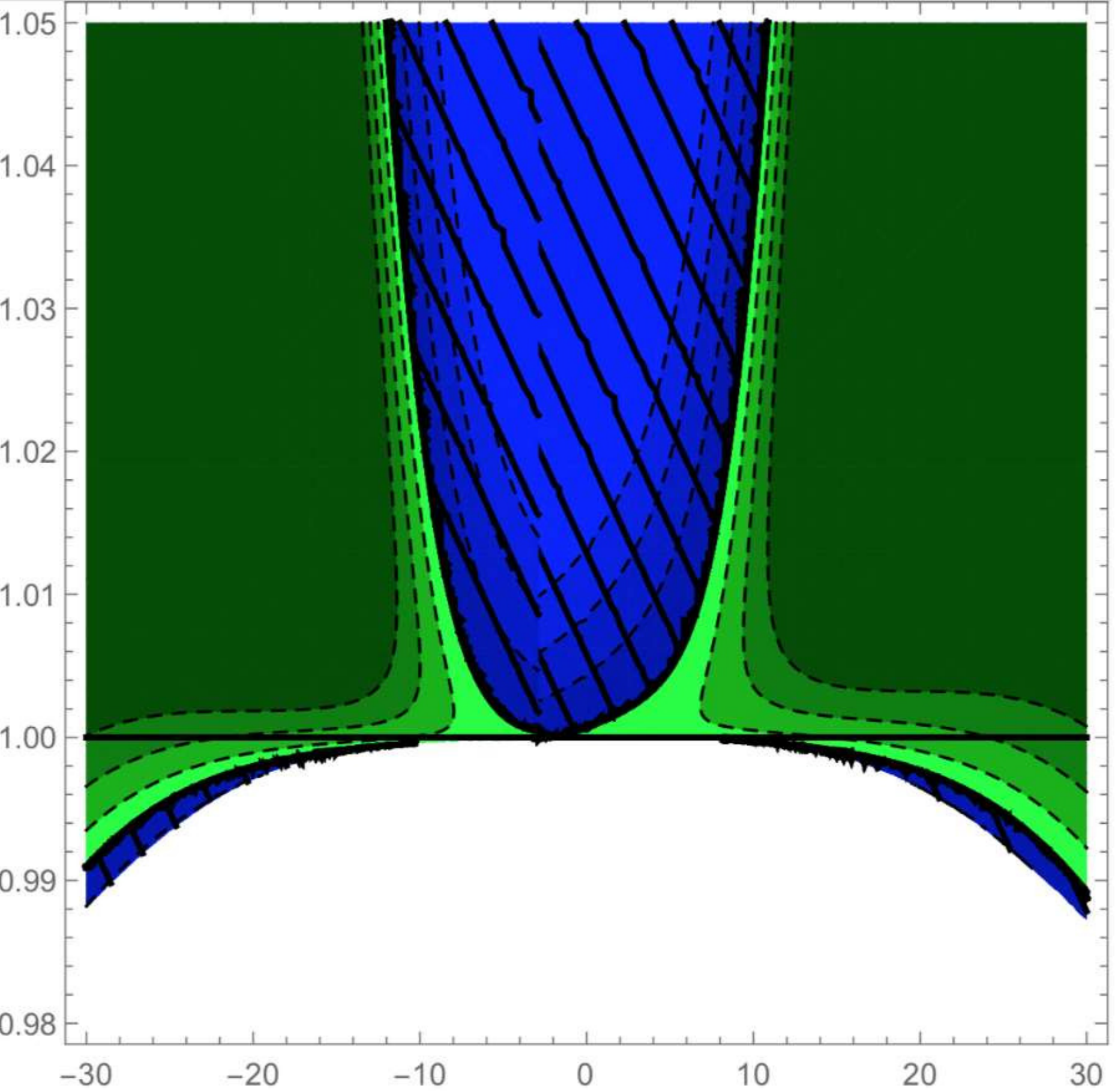}}
\hspace{0.1cm}$\pi$\hspace{2.8cm}\hspace{-2.2cm}
{\includegraphics[width=0.08\linewidth] {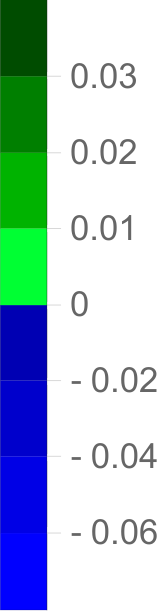}}
\caption{[color online] Scalar sound speed squared
$c_\mathcal{S}^2(\pi,\dot{\pi})$ in the
  phase space $(\pi,\dot{\pi})$ for the original model with 
  bouncing solution~\cite{bounceI}.
 Dashed lines are lines of constant 
 $c_\mathcal{S}^2(\pi,\dot{\pi})$. 
As in Fig.~\ref{fig:tensor_old1}, there are no solutions 
in the white region.
Negative values of  $c_\mathcal{S}^2(\pi,\dot{\pi})$ 
(hatched region) mean gradient instability.
For small $\pi$, the solution manages to safely
pass between the
white forbidden region and the hatched one with $c_\mathcal{S}^2<0$.
} 
\label{fig:scalar_old1}
\end{center}\end{figure}

Let us now turn to the propagation speeds of perturbations
about the homogeneous backgrounds in the vicinity
of the original solution.
We parametrize the phase space of homogeneous solutions
by $\pi$ and $\dot{\pi}$, then the original solution is a line
$\dot{\pi}=1$, $\pi \in (-\infty, +\infty)$, and its vicinity
is a strip along this line.
The background equation
\eqref{eq:dg00} is used to determine $H$ in terms of $\pi$ and $\dot{\pi}$.
Note that eq.~\eqref{eq:dg00} is a quadratic equation for $H$,
so there are regions in phase space $(\pi, \dot{\pi})$ where the solution
does not exist. The values of $\dot{H}$ and $\ddot{\pi}$ are obtained,
for given  $\pi$, $\dot{\pi}$, from eq.~\eqref{eq:dgii} and
time derivative of eq.~\eqref{eq:dg00}. We plug 
$H$, $\dot{H}$ and $\ddot{\pi}$
in eqs.~\eqref{eq:GS_setup} -- \eqref{eq:sound_speeds}
and obtain the desired propagation speeds
$c_\mathcal{T}^2(\pi, \dot{\pi})$ and $c_\mathcal{S}^2(\pi, \dot{\pi})$.

The explicit expressions for $c_\mathcal{T}^2(\pi, \dot{\pi})$
and $c_\mathcal{S}^2(\pi, \dot{\pi})$ are cumbersome, and we do 
not give them here. 
The speed squared
of gravitational waves  $c_\mathcal{T}^2(\pi, \dot{\pi})$
as a 
function of phase space points is shown in Fig.~\ref{fig:tensor_old1}.
We see that there are regions (with $\dot{\pi}<1$)
in the immediate vicinity of the bouncing solution
$\dot{\pi}=1$, where the propagation of gravitational waves is
superluminal. Thus, the original model of Ref~\cite{bounceI} is
unsatisfactory, as it does not
obey the requirement of the absence of superluminality.

We present for completeness the sound speed of scalar perturbations
in Fig.~\ref{fig:scalar_old1}. There is no superluminality
in a vicinity of the original
solution $\dot{\pi}=1$.

Thus,
choosing
the Ansatz \eqref{eq:lagr_series} and requiring
$\mathcal{{G}_T} =\mathcal{{F}_T} = 1$ on the solution $\pi(t) = t$
is not a good idea.
In the next Section 
we discuss ways of avoiding
{this
superluminality problem with gravitational waves}
and 
suggest a modified version of the stable
bouncing solution, which
has in its vicinity
safely subluminal perturbations in both tensor and scalar sector.

\section{Subluminal  bouncing solution}
\label{sec:solution}
Superficially, one can think of two  approaches to avoid
 superluminality in the tensor sector. The first one amounts to
changing the Ansatz~\eqref{F4} for the Lagrangian so that the
function $F_4$ involves a quadratic contribution $f_{42}(\pi)\cdot X^2$.
Then it might be possible to
still take $\mathcal{G_T} = \mathcal{F_T} = 1$ on the solution $X=1$, while 
making $\mathcal{G_T} < 1$, and hence
$c_\mathcal{T}^2 < 1$, in a vicinity of the
solution by adjusting the new function $f_{42}(\pi)$.
This is conceivable, since $f_{42}(\pi)$ 
contributes to $\mathcal{G_T}$ but not to $\mathcal{F_T}$, 
see eqs.~\eqref{eq:GT_coeff_setup}. It appears, however, that
this approach requires 
fine-tuning of the functions in the Ansatz.
Moreover, according to
eqs.~\eqref{eq:GS_setup} and~\eqref{eq:FS_setup}, both
$\mathcal{G_S}$ and $\mathcal{F_S}$ would involve $f_{42}(\pi)$,
hence, one would have to take care of the subluminal propagation 
and absence of instabilities in 
the scalar sector (note that originally the scalar modes were 
automatically subluminal). 

Another way to deal with the superluminality threat is to
abandon the requirement that $c_\mathcal{T}^2 =1$ on the bouncing
solution. Let us follow this route, i.e.
construct a model with the stable bouncing solution that has
 $c_\mathcal{T}^2 < 1$. Then a small deviation from the
solution
 will not have superluminal excitations too.
At the same time we make sure that
$c_\mathcal{T}^2(t)|_{t\to\pm\infty} \to 1$ in accordance with
GR form of the asymptotics. Needless to say, our construction
involves the Lagrangian functions different from those in
Sec.~\ref{sec:superluminal}.

\begin{figure}[h]
\center{\hspace{-1cm}\includegraphics[width=0.4\linewidth]{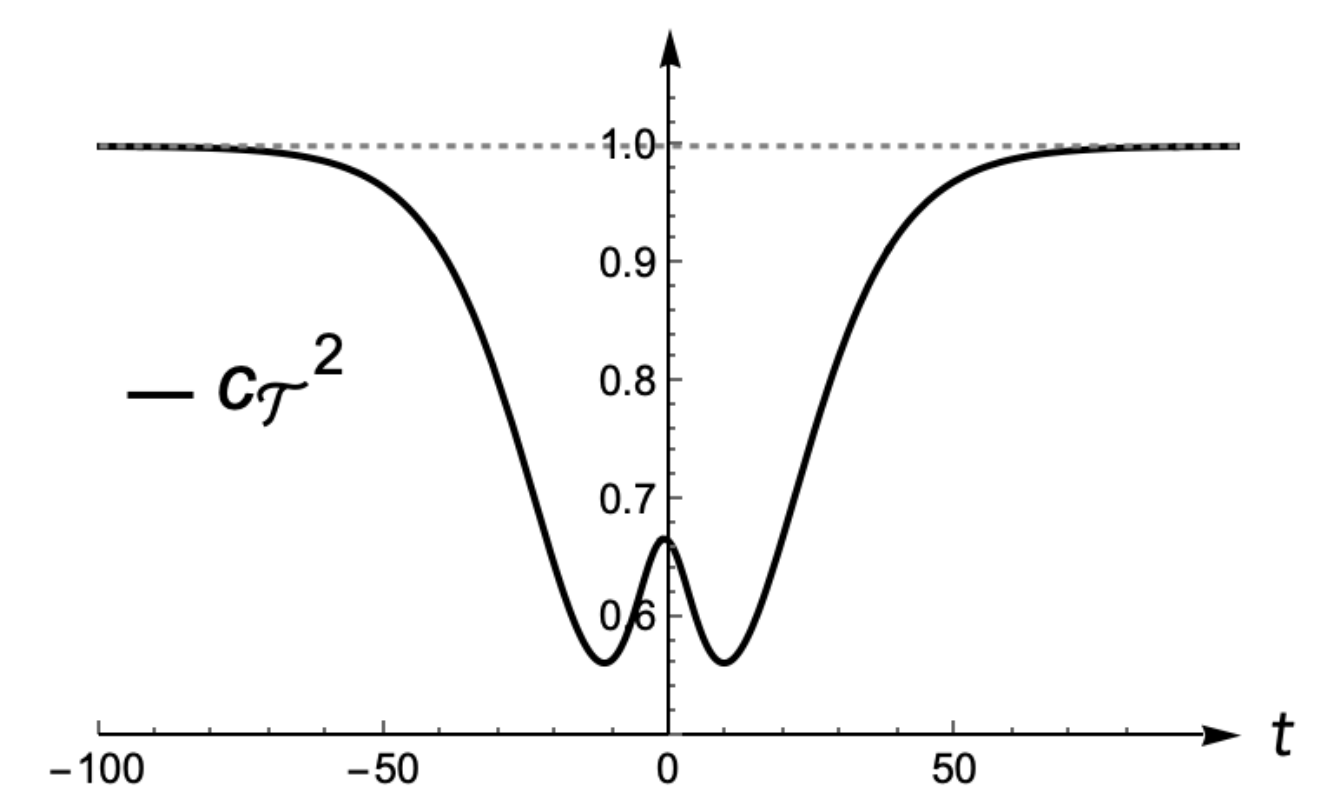}\hspace{0.6cm}\includegraphics[width=0.4\linewidth]{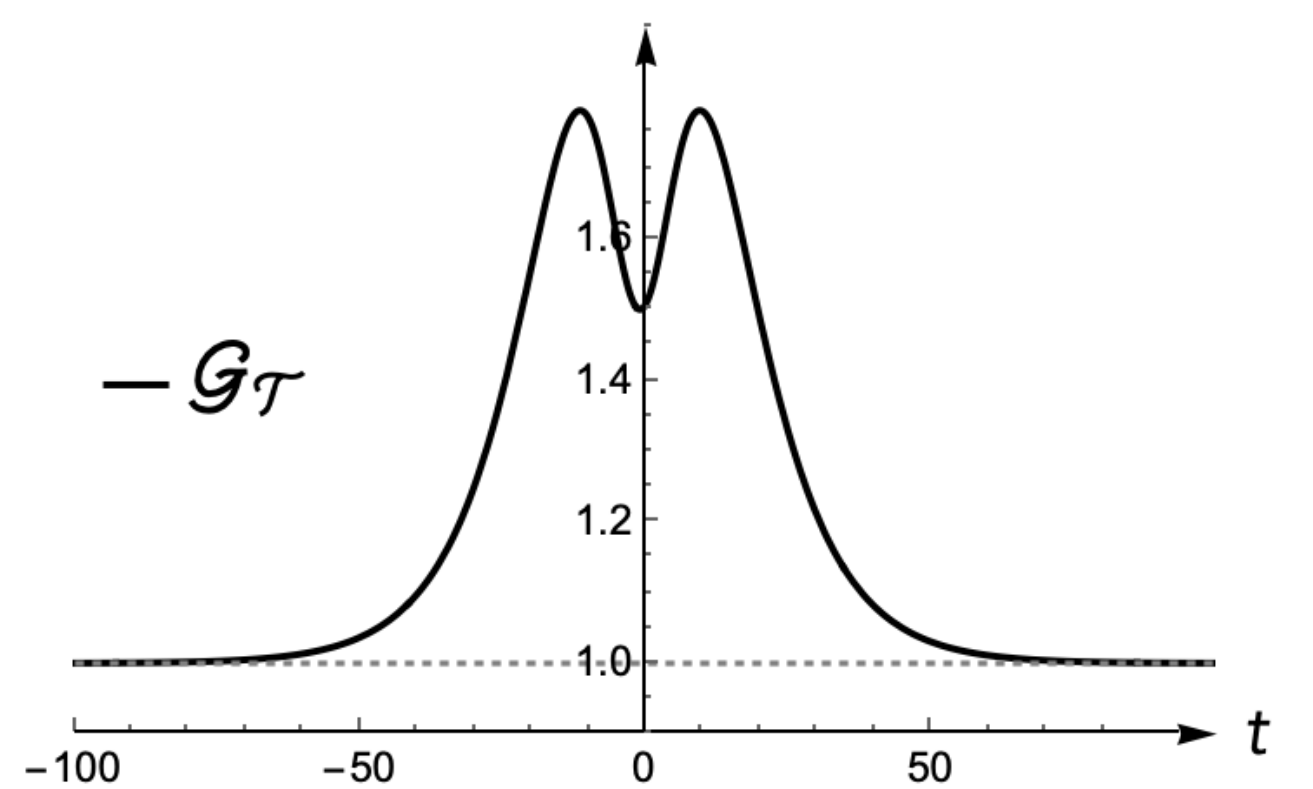} }
\caption{The speed squared of tensor perturbations
  is non-negative, smaller than 1
  for all times and asymptotically tends to
  1 in both infinite past and future.
  Right panel shows the behaviour of the coefficient $\mathcal{G_T}$
in eq.~\eqref{eq:newGtt} with $u=1/10$, $w=1$ and $\tau=10$.}
\label{fig:ct_Gt}
\end{figure}

\begin{figure}[h!]\begin{center}\hspace{-1cm}
{\includegraphics[width=0.5\linewidth]{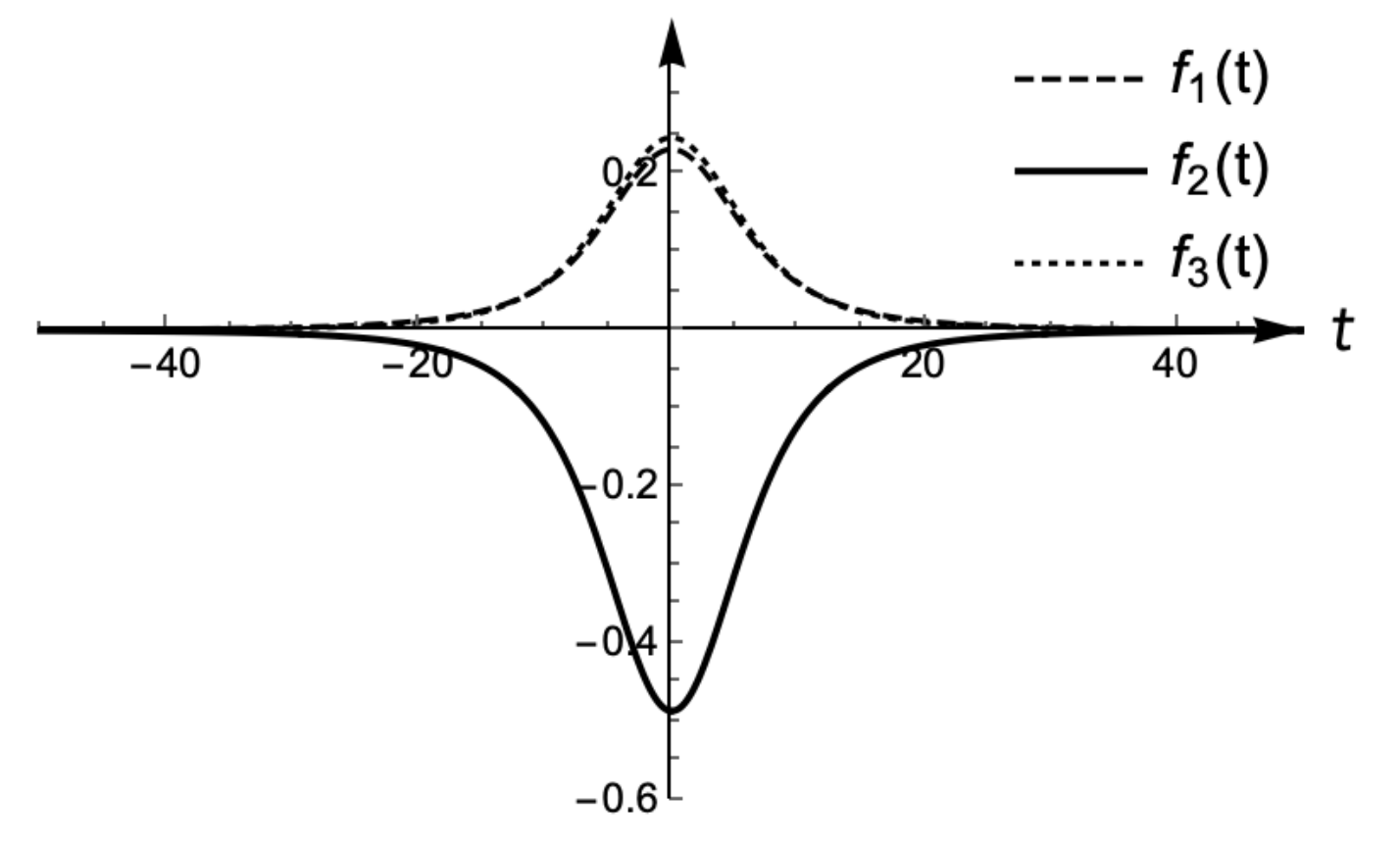}}\hspace{2.8cm}\hspace{-3cm}
{\includegraphics[width=0.5\linewidth] {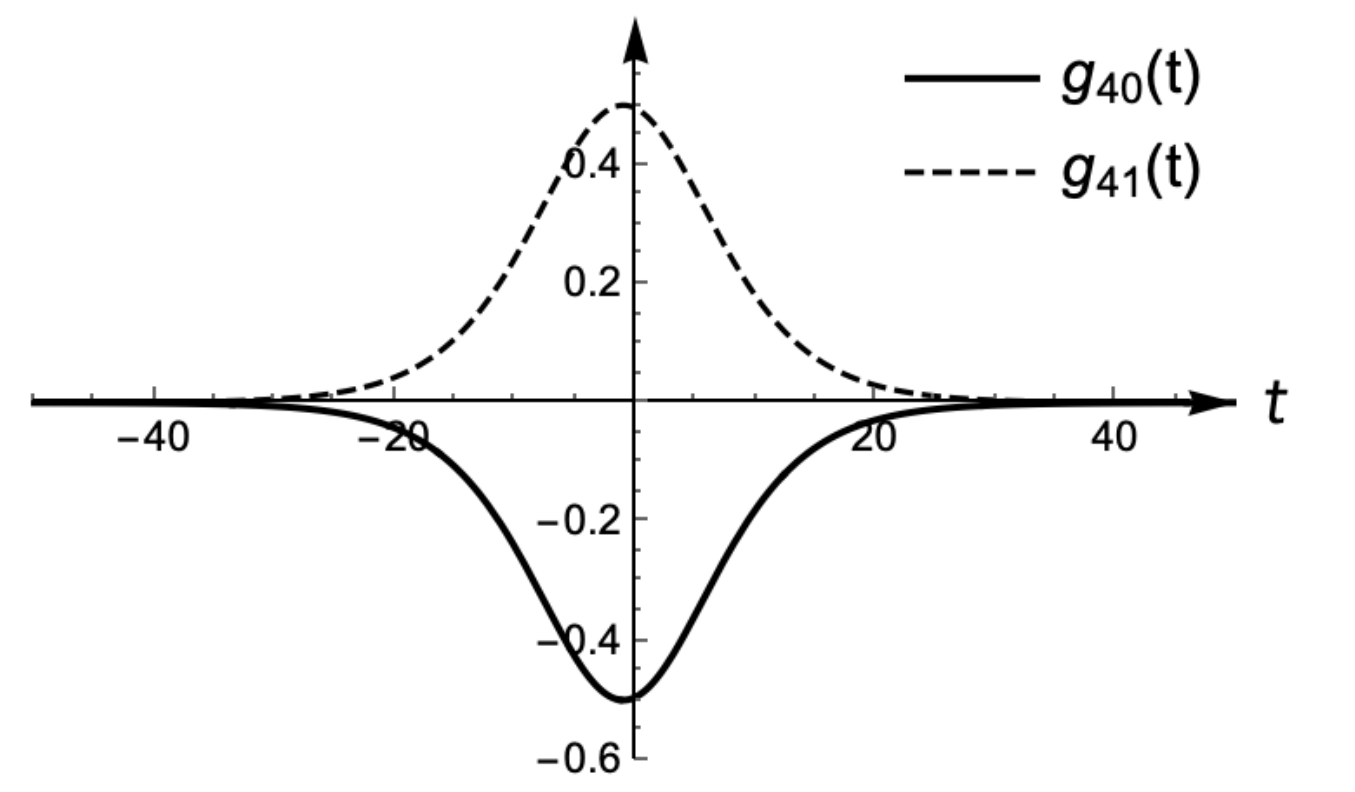}}

\vspace{-0.1cm}
{\includegraphics[width=0.5\linewidth]
{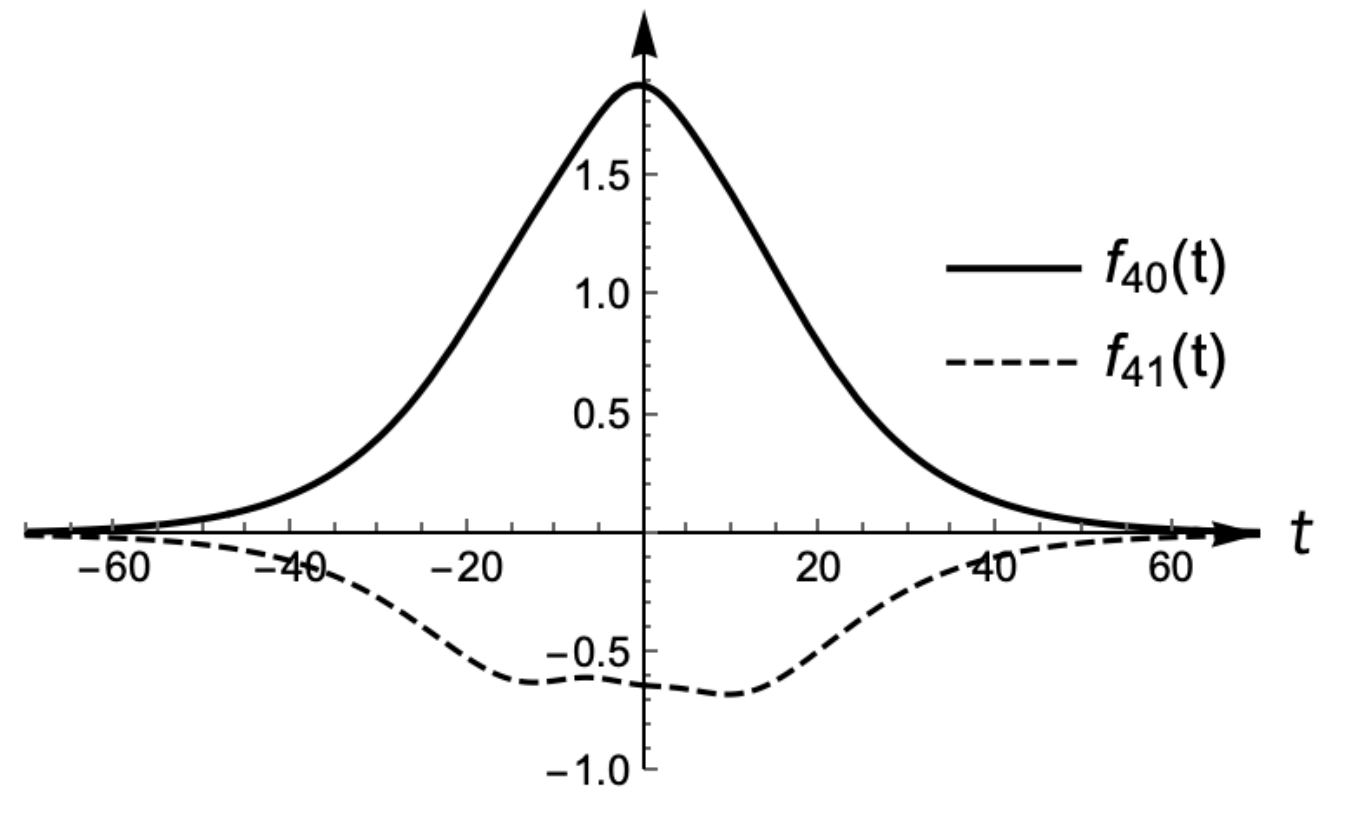}}\hspace{2.cm}
\caption{The Lagrangian functions $f_1(t)$, $f_2(t)$, $f_3(t)$, $g_{40}(t)$, $g_{41}(t)$,
$f_{40}(t)$ and $f_{41}(t)$, with the following choice of the parameters involved in the analytical expressions (see Appendix B): $u=1/10$, $w=1$ and $\tau = 10$. This choice guarantees that the bouncing solution is not fine-tuned and its duration safely exceeds the Planck time. Note that the functions $f_1(t)$ and $f_3(t)$ almost coincide for the chosen values of parameters.} 
\label{fig:LagrangianFunctions}
\end{center}\end{figure}

\begin{figure}[h]
\center{\hspace{-1cm}\includegraphics[width=0.4\linewidth]{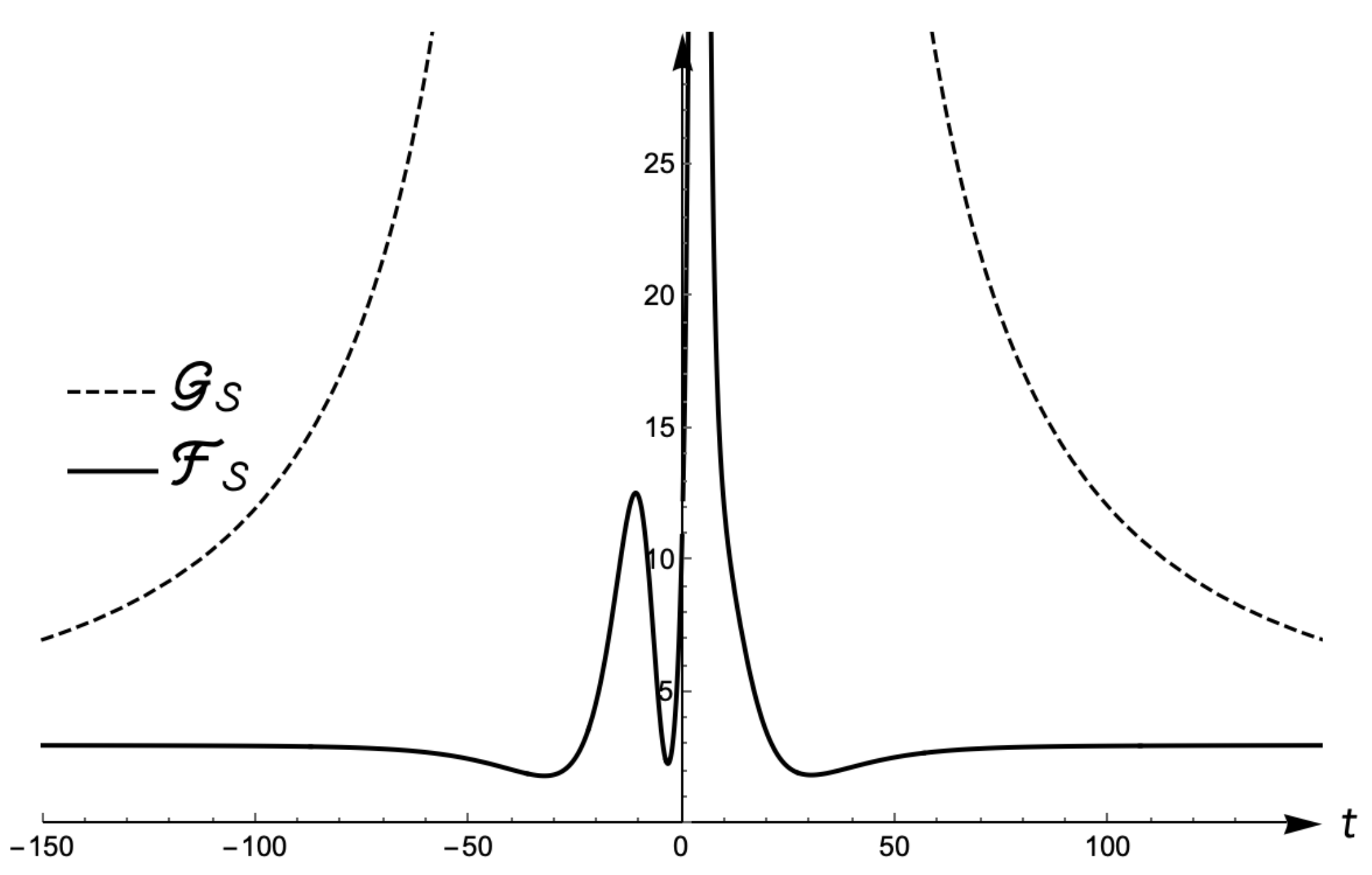}\hspace{5cm}\includegraphics[width=0.4\linewidth]{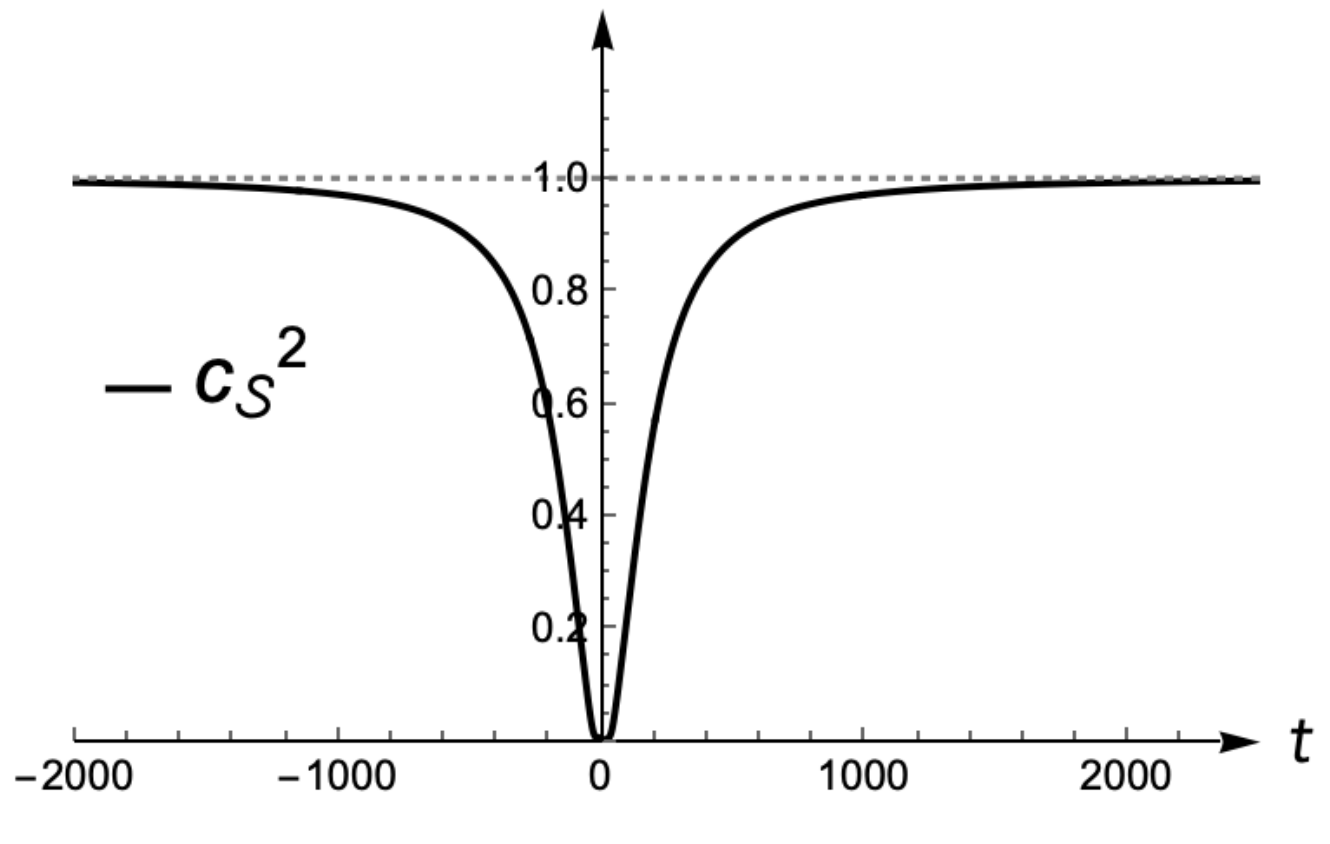} 
\hspace{0.5cm}\includegraphics[width=0.4\linewidth]{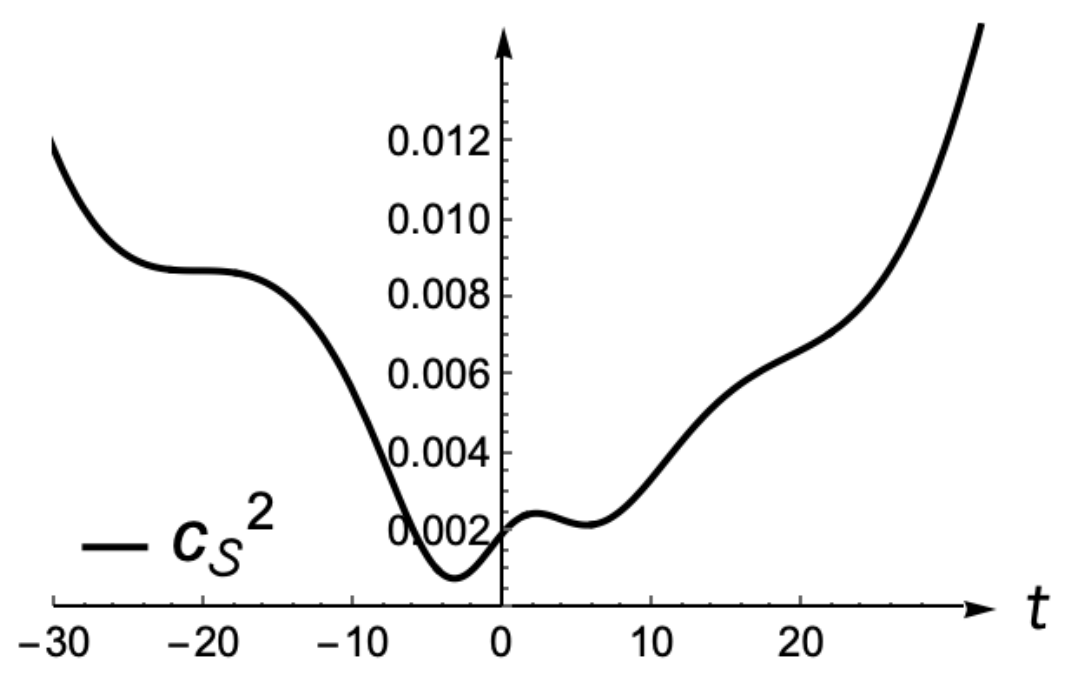}}
\put(-342,30){\line(1,0){30}}
\put(-342,00){\line(1,0){30}}
\put(-342,30){\line(0,-1){30}}
\put(-312,00){\line(0,1){30}}
\put(-200,0){\line(1,0){200}}
\put(-200,125){\line(1,0){200}}
\put(-200,0){\line(0,1){125}}
\put(0,125){\line(0,-1){125}}
\put(-342,30){\line(3,2){142}}
\put(-312,0){\line(1,0){120}}

\caption{The coefficients $\mathcal{G_S}$ and $\mathcal{F_S}$ 
(top panel) and $c_\mathcal{S}^2$ (bottom panels). The sound speed squared for the scalar perturbations is non-negative for all times and asymptotically tends to 1 in both infinite past and future.}
\label{fig:gsfs_cs}
\end{figure}

\begin{figure}[h!]\begin{center}\hspace{-1cm}
\put(0,300){$\dot{\pi}$}
{\includegraphics[width=0.6\linewidth]{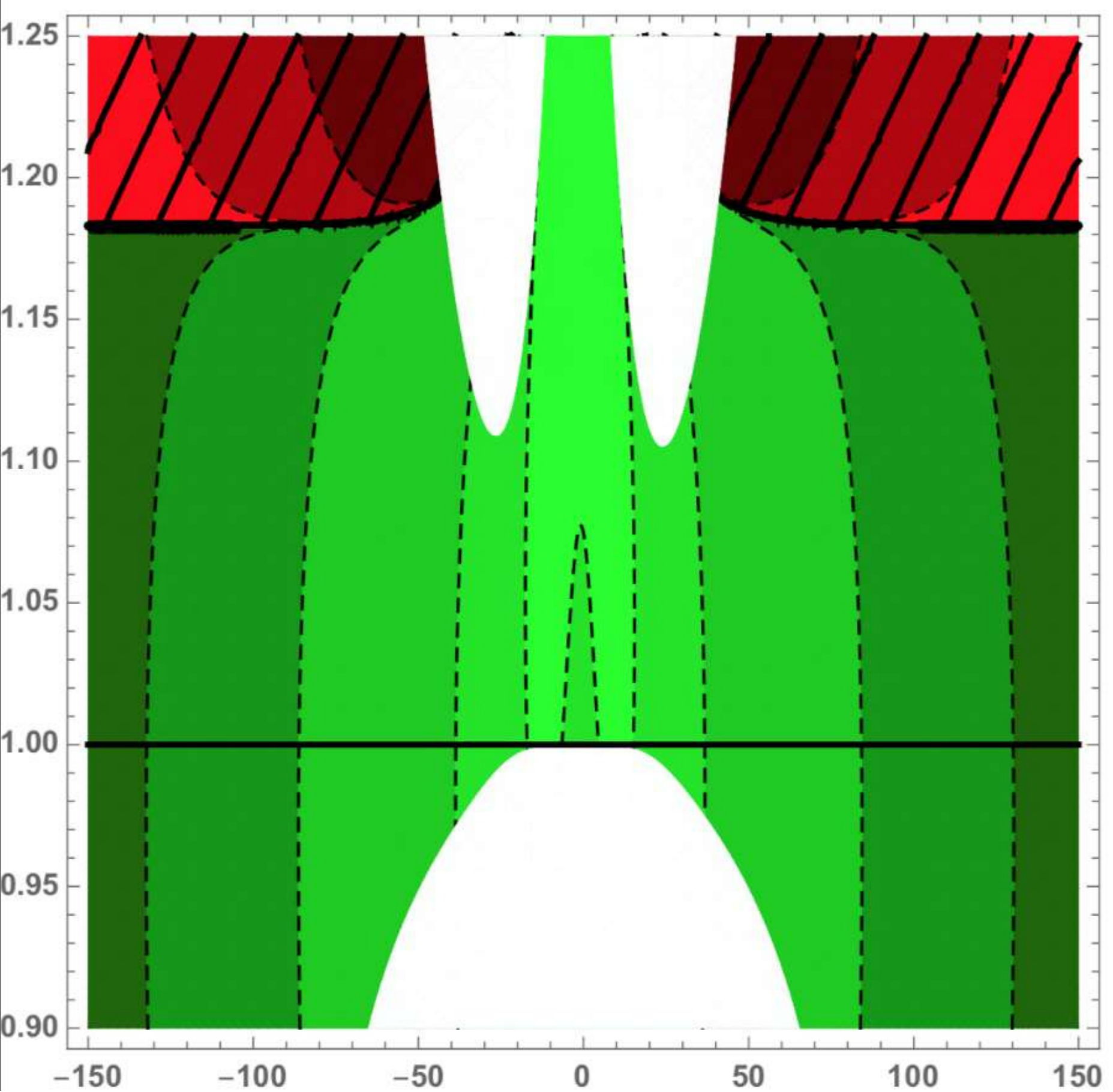}}
\hspace{0.1cm}$\pi$\hspace{2.8cm}\hspace{-2.5cm}
{\includegraphics[width=0.1\linewidth] {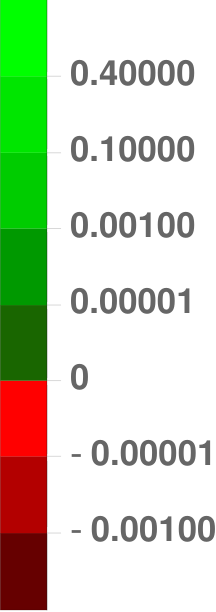}}
\caption{[color online] Variance
  $(1 - c_\mathcal{T}^2(\pi,\dot{\pi}))$ of the speed squared of tensor
  modes in the
  phase space $(\pi,\dot{\pi})$ for the new 
bouncing solution. Dashed lines are lines of constant 
$(1 - c_\mathcal{T}^2(\pi,\dot{\pi}))$.
A thick line in the upper part denotes the boundary
of the (hatched) superluminal region. The black 
horizontal line at $\dot{\pi}=1$
corresponds to the new bouncing solution.
The plot shows that there are no superluminal tensor
perturbations in a vicinity of the bouncing solution.
}\label{fig:tensor_new1}
\end{center}\end{figure}
\begin{figure}[h!]\begin{center}\hspace{-0.6cm}
\put(0,200){\footnotesize $\dot{\pi}$}
\put(275,200){\footnotesize $\dot{\pi}$}
\put(200,0){\footnotesize $\pi$}
\put(465,0){\footnotesize $\pi$}
{\includegraphics[width=0.4\linewidth]{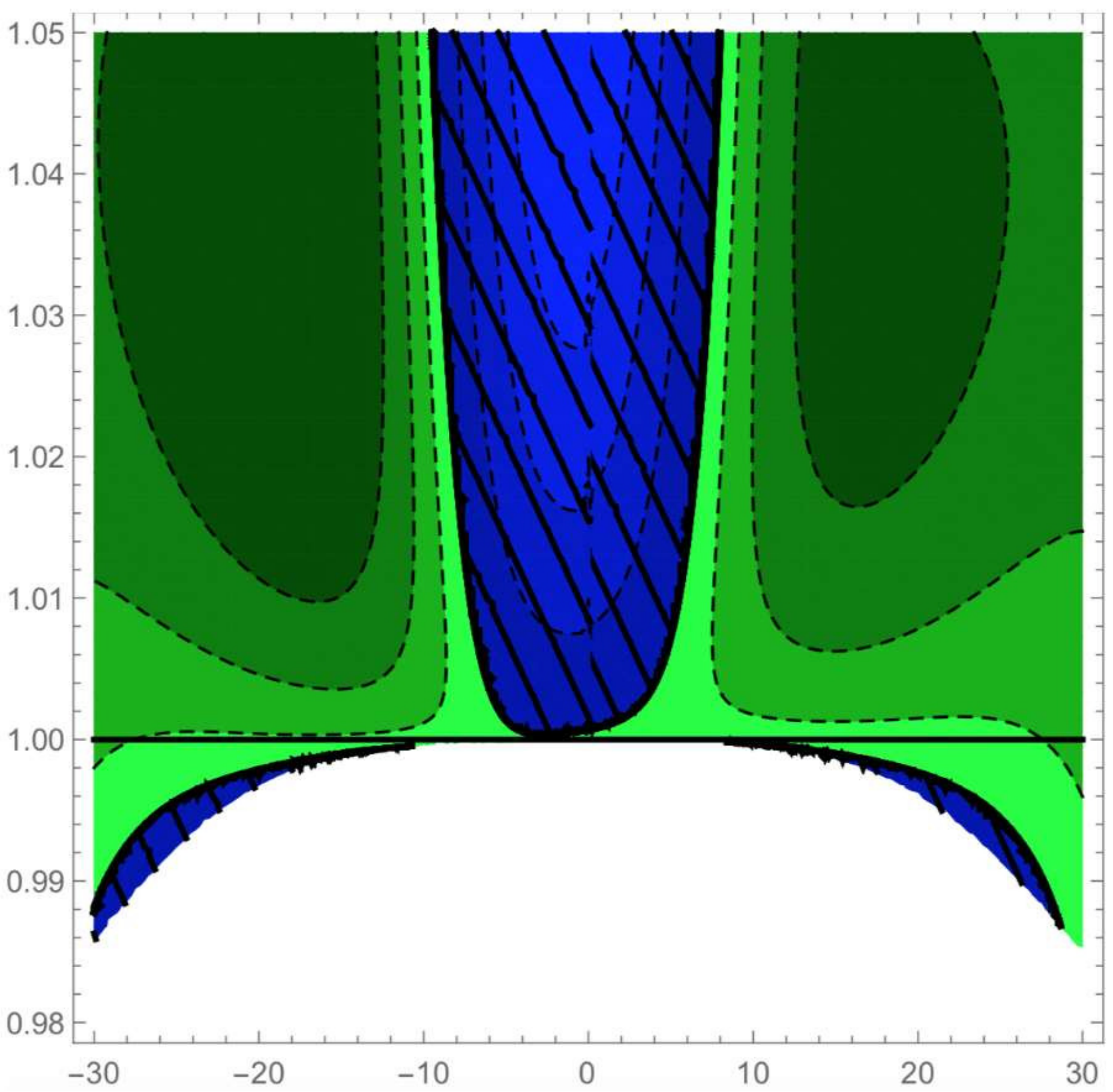}}\hspace{2.8cm}\hspace{-2.6cm}
{\includegraphics[width=0.07\linewidth]{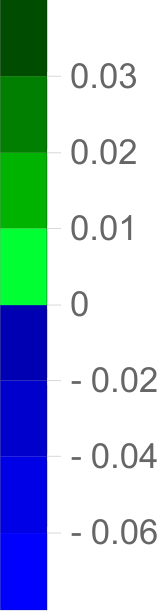}}
\hspace{0.5cm}
{\includegraphics[width=0.4\linewidth]{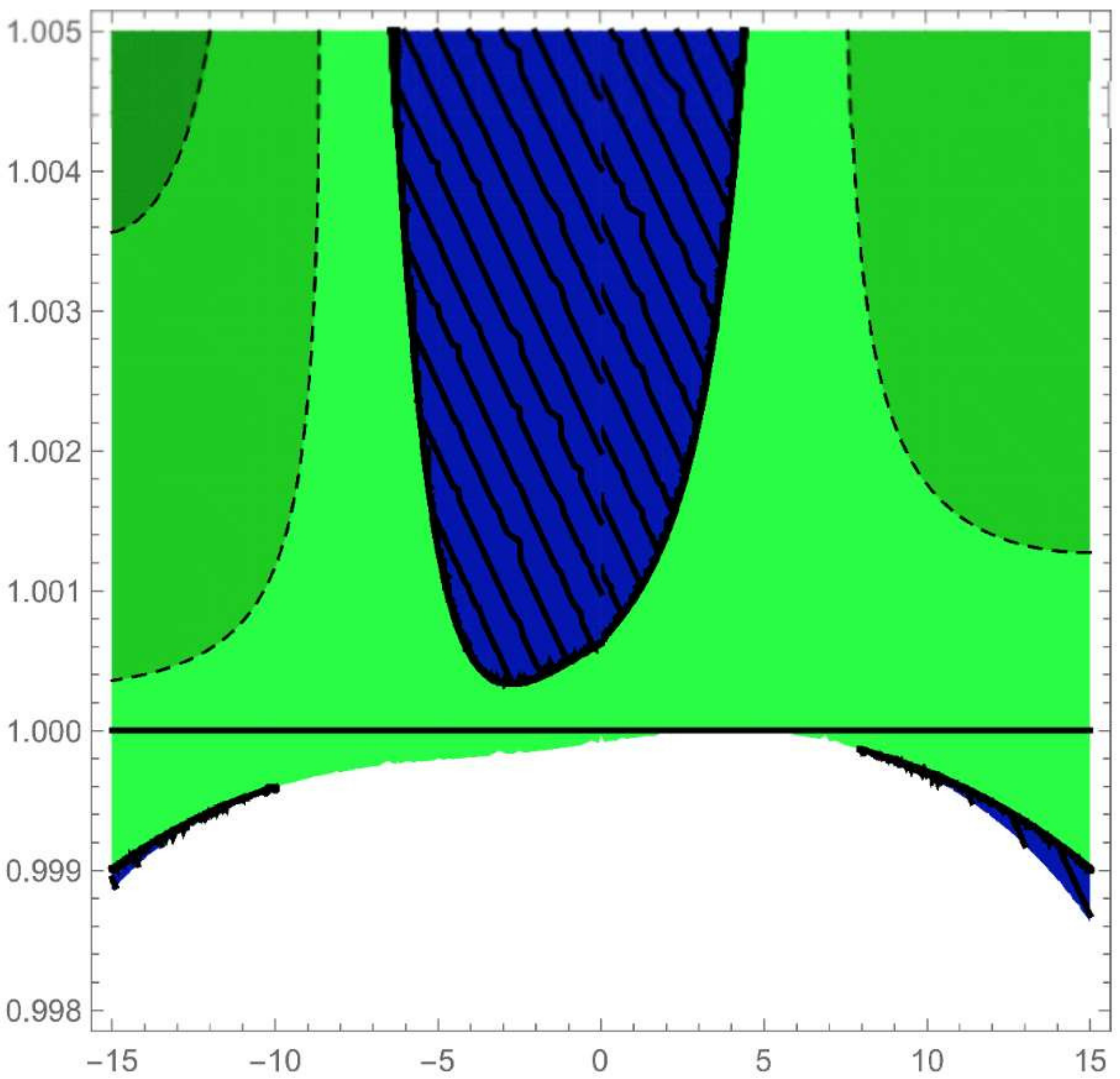}}\hspace{2.8cm}\hspace{-2.6cm}
\includegraphics[width=0.08\linewidth]{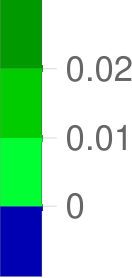}
\caption{[color online] Scalar sound speed squared
$c_\mathcal{S}^2(\pi,\dot{\pi})$ in the
  phase space $(\pi,\dot{\pi})$
  for the new bouncing solution.
  The black horizontal line corresponds to the solution
  and is again safely away from both
  the white zone with no solutions and the pathological hatched zone.
  Right panel is a blow up of the bounce region.
} \label{fig:scalar_new1}
\end{center}\end{figure}

We still take the Hubble parameter in the form~\eqref{eq:Hubble}, a
rolling scalar field $\pi(t)=t$ and now choose, instead
of~\eqref{eq:lagr_series},
the 
Ansatz (this is a matter of convenience)
\begin{subequations}
\label{eq:lagr_series-bis}
\begin{align}
\label{F-bis}
& F(\pi, X) = f_1(\pi)\cdot X + f_2(\pi)\cdot X^2 + f_3(\pi)\cdot X^3, \\
\label{G4-bis}
& G_4(\pi, X) = \frac12 + g_{40}(\pi) + g_{41}(\pi) \cdot X,\\
\label{F4-bis}
& F_4(\pi, X) = f_{40}(\pi) + f_{41}(\pi) \cdot X.
\end{align}
\end{subequations}
In contrast to our previous construction we keep only
$\mathcal{F_T} = 1$, while taking
\be
\label{eq:newGtt}
\mathcal{G_T} = 1 + \frac{5w}{2} \;\mbox{sech}\left(\frac{t}{\tau}+u\right)
- 2\; \mbox{sech}^2\left(\frac{t}{\tau}+u\right),
\ee
with an arbitrary parameters $u$ and $w$ (we use $t$ and $\pi$
interchangeably as long as the bouncing solution is discussed).
The choice in eq.~\eqref{eq:newGtt} 
makes $c_\mathcal{T}^2(\pi) = \mathcal{F_T}/\mathcal{G_T}$
behave as shown in Fig.~\ref{fig:ct_Gt}, where we also plot 
$\mathcal{G_T}$. 
We skip the further reconstruction steps, since they are
basically the same as in
Ref.~\cite{bounceI}, and give the results only.

The reconstructed Lagrangian functions~\eqref{eq:lagr_series-bis}
are shown in Fig.~\ref{fig:LagrangianFunctions},
while the analytical expressions are given in Appendix B.
The asymptotic behavior of the functions as $t \to \pm \infty$
is as follows:
\begin{equation}
\label{eq:lagrangian_asymptotics}
f_1(t) = \frac{1}{3 t^2},  \;\;f_0(t) = f_2(t) \propto \frac{1}{t^4}, \;\; g_{40}(t) = g_{41}(t)  \propto e^{- 2|t|/\tau}, \;\; f_{40}(t) = f_{41}(t)\propto e^{-|t|/\tau},
\end{equation}
which indeed corresponds to the asymptotic Lagrangian 
(see eqs.~\eqref{eq:lagr_series-bis} and~\eqref{eq:lagrangian})
\be
\label{eq:lagrangian_asymp}
\mathcal{L}_{t\to\pm\infty} = -\frac12 R + \dfrac{X}{3\pi^2},
\ee
and describes GR and a conventional massless scalar field
$\phi = \sqrt{\frac23} \log(\pi)$
in both distant past and future.
We plot corresponding $\mathcal{G_S}$, $\mathcal{F_S}$ and
$c_\mathcal{S}^2$ as functions of $\pi=t$ in Fig.~\ref{fig:gsfs_cs} 
in order to illustrate
that the new bouncing solution
does not involve ghost
and gradient instabilities as well as superluminal modes in the 
scalar sector at all times.

Figs.~\ref{fig:tensor_new1} and \ref{fig:scalar_new1} illustrate
that the superluminality issue  is indeed
resolved 
{in the modified model:}
the new bouncing solution is safely far away from 
the superluminal regions in $(\pi,\dot{\pi})$ plane for both tensor 
and scalar modes.
A peculiar property of our set up is that the speed of tensor
perturbations tends to 1 as $t\to\pm\infty$, and yet
there is always a finite gap in the phase space between our solution
$\dot{\pi}=1$ and the line $c_\mathcal{T}^2(\pi,\dot{\pi})=1$, 
see Fig.~\ref{fig:tensor_new1}. 
Let us note that as it stands,
the new solution is not separated from
the superluminal domains by any kind of forbidden area, hence, these
domains are in principle reacheable.
However, we recall that
the
reconstructed Lagrangian~\eqref{eq:lagr_series-bis} is valid only in a
vicinity of the solution with $X=1$. Therefore, the would-be superluminal
domains are away from the region of validity of our analysis,
so we consider our construction healthy
{at least in the absence of additional matter.}

\begin{figure}[h!]\begin{center}\hspace{-1cm}
\put(0,225){\footnotesize $\dot{\pi}$}
\put(260,225){\footnotesize $\dot{\pi}$}
\put(230,0){\footnotesize $\pi$}
\put(485,0){\footnotesize $\pi$}
{\includegraphics[width=0.45\linewidth]{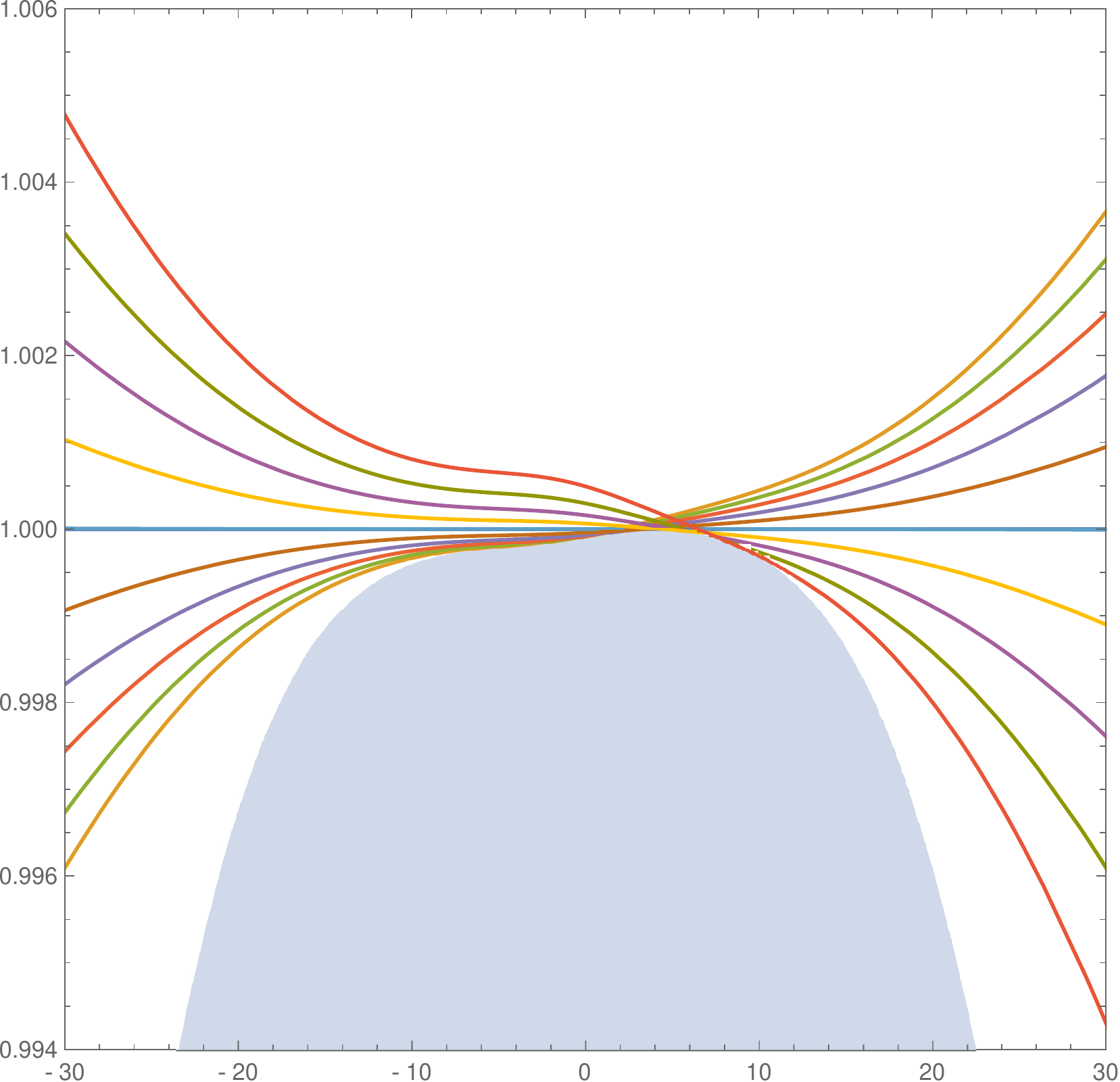}}
\hspace{0.7cm}
{
\includegraphics[width=0.45\linewidth]{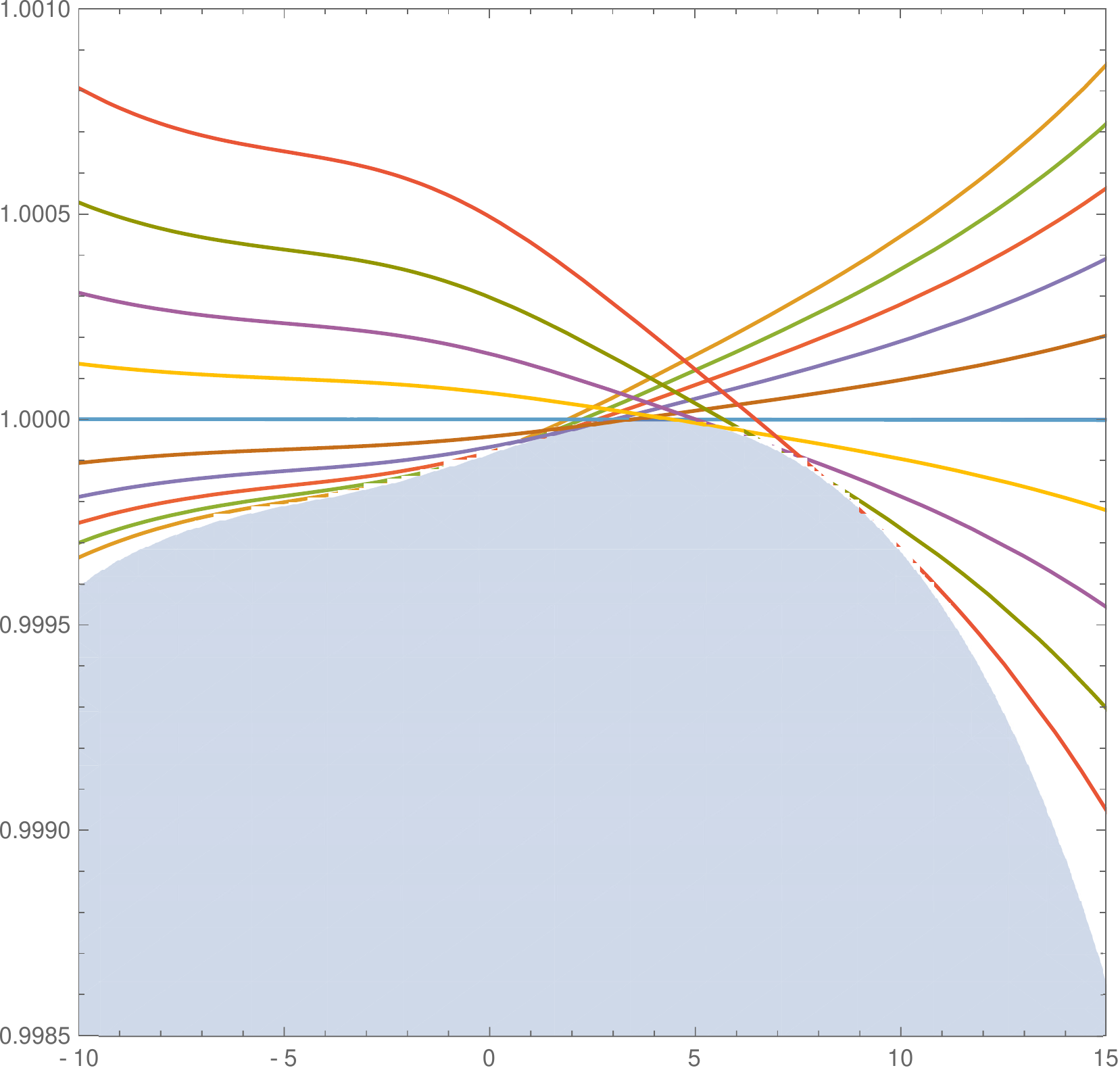}}
\caption{[color online] Bouncing trajectories with different initial conditions
  in $(\pi,\dot{\pi})$ plane.
There are no solutions inside
the shaded region, while the trajectories
touch the boundary of this region and exhibit
required $\gamma$-crossing. 
The right panel is a blow up of the bounce region, with the same solutions
as in the left panel.
} \label{fig:gamma_crossing}
\end{center}\end{figure}

To end up this section, we discuss whether our new bouncing solution
requires strong fine-tuning of the initial data.
To this end, we stick to
the reconstructed Lagrangian functions~\eqref{eq:lagr_series-bis} as
given in Appendix B,
and solve the system of background gravitational 
equations~\eqref{eq:Einstein} for $H$ and
$\pi$  for
different initial conditions. The resulting trajectories in the
phase space 
$(\pi,\dot{\pi})$ are shown in
Fig.~\ref{fig:gamma_crossing}. 
The shaded region
in both figures
(which is
of the same nature as the white regions in Figs.~\ref{fig:tensor_old1},
~\ref{fig:scalar_old1},~\ref{fig:tensor_new1} and~\ref{fig:scalar_new1})
is a forbidden domain, 
where eq.~\eqref{eq:dg00}, viewed as  the quadratic equation
for $H$, has negative
discriminant.
As pointed out in Ref.~\cite{Unbraiding} and in 
{Sec.~\ref{sec:superluminal}},
zero  discriminant, i.e. the transition between
the branches of solution for the Hubble parameter, occurs at
$\gamma$-crossing. Since we have required that
the beyond Horndeski theory~\eqref{eq:lagr_series-bis}  
tends to
GR + massless scalar field long before and long after the 
bouncing epoch, the $\gamma$-crossing has to take place at some 
moment of time. Hence, 
the bouncing trajectories have to touch 
the boundary of the 
forbidden domain, which is indeed the case in Fig.~\ref{fig:gamma_crossing}.
{In particular,}
  before touching the boundary of the forbidden domain (on the left
  of $\gamma$-crossing)
  each of these trajectories corresponds to the smaller root $H_-$
  of eq.~\eqref{eq:dg00}, 
  while after the $\gamma$-crossing
  the Hubble parameter is given by the larger root
  $H_+$ of eq.~\eqref{eq:dg00}.
We have checked that every trajectory in Fig.~\ref{fig:gamma_crossing}
describes healthy bouncing solution. 
Since there is a whole set of these trajectories with
different initial conditions, we conclude
that
no fine-tuning is involved in the solution construction.


\section{{Adding} extra matter}
\label{sec:extra_matter}
In the previous section we have demonstrated that
one {can}
arrange the beyond Horndeski Lagrangian in such 
a way that both tensor and scalar perturbations 
about the bouncing solution 
are safely subluminal. However,
{nothing appears to prevent one
{from adding}
other matter components, e.g. radiation or dust, to}
the cosmological setup. 
Thus, it is legitimate to ask how these extra matter components affect 
stability and propagation speed of 
perturbations
in a cosmological setting of beyond Horndeski theory, 
and our model in particular.
{Importantly, one may wonder whether
  additional matter makes our model of Sec.~\ref{sec:solution}
    problematic
by inducing} superluminality 
in the same way as it happened with the subluminal Genesis scenario in
the cubic Horndeski subclass~\cite{MatMat}.
{We recall in this regard that
{we have not defined} 
  our Lagrangian away from the vicinity
  of the region with $X=1$, so we limit our
  discussion to the region of the phase space near $\dot{\pi}=1$.}

The case of (beyond) Horndeski theory + {additional}
matter was studied in various contexts,
see, e.g. Refs.~\cite{speed1,MatMat,speed2,speed3,KobaRev}.
In this section we {first derive}
 expressions for stability conditions 
 and propagation speeds
 for perturbations
{about cosmological solutions}
  in {general}
  beyond Horndeski {theory}
with an
{additional ideal fluid} component 
\footnote{Our notations are similar to those
utilized in Horndeski case with additional k-essence 
in Ref.~\cite{KobaRev}.}.
  {Then we} specify to the model of Sec.~\ref{sec:solution} to figure out
  whether there exist stable points in the phase space near $\dot{\pi}=1$
  featuring
superluminal propagation of perturbations.

Let us consider a model where along with the scalar
field of beyond Horndeski type there is 
{perfect fluid with the standard  stress-energy tensor}
\bea
\label{eq:SET}
& T_{\mu\nu} = (p+\rho)u_{\mu}u_{\nu}- p\,g_{\mu\nu},
\eea
where $\rho$,{
$p$ and $u_{\mu}$ are
energy density,  pressure and 4-velocity of the fluid,
respectively}
($g^{\mu\nu} u_{\mu} u_{\nu} = 1$). 

In the presence of
extra fluid, the {gravitational
equations
for
spatially flat FLRW background in general
beyond Horndeski theory read (cf. eqs~\eqref{eq:dg00}-\eqref{eq:dgii})}
\begin{subequations}
\label{eq:Einstein_rad}
\begin{align}
\label{eq:dg00_rad}
\delta g^{00}: \;\;
&F-2F_XX-6HK_XX\dot{\pi}+K_{\pi}X+6H^2G_4
+6HG_{4\pi}\dot{\pi}-24H^2X(G_{4X}+G_{4XX}X)
\\\nonumber&+12HG_{4\pi X}X\dot{\pi}
-2H^3X\dot{\pi}(5G_{5X}+2G_{5XX}X)+3H^2X(3G_{5\pi}+2G_{5\pi X}X)
\\\nonumber&+6H^2X^2(5F_4+2F_{4X}X)
+6H^3X^2\dot{\pi}(7F_5+2F_{5X}X)- \rho= 0,\\
\label{eq:dgii_rad}
\\\nonumber
\delta g^{ii}: \;\;
&F-X(2K_X\ddot{\pi}+K_\pi)+2(3H^2+2\dot{H})G_4-12H^2G_{4X}X
-8\dot{H}G_{4X}X-8HG_{4X}\ddot{\pi}\dot{\pi}
\\\nonumber&-16HG_{4XX}X\ddot{\pi}\dot{\pi}
+2(\ddot{\pi}+2H\dot{\pi})G_{4\pi}+4XG_{4\pi X}(\ddot{\pi}-2H\dot{\pi})+2XG_{4\pi\pi}
\\\nonumber&-2XG_{5X}(2H^3\dot{\pi}+2H\dot{H}\dot{\pi}+3H^2\ddot{\pi})+G_{5\pi}(3H^2X+2\dot{H}X+4H\ddot{\pi}\dot{\pi})-4H^2G_{5XX}X^2\ddot{\pi}
\\\nonumber&+2HG_{5\pi X}X(2\ddot{\pi}\dot{\pi}-HX)
+2HG_{5\pi\pi}X\dot{\pi}+2F_4X(3H^2X+2\dot{H}X+8H\ddot{\pi}\dot{\pi})
\\\nonumber&+8HF_{4X}X^2\ddot{\pi}\dot{\pi}+4HF_{4\pi}X^2\dot{\pi}+6HF_5X^2(2H^2\dot{\pi}+2\dot{H}\dot{\pi}+5H\ddot{\pi})
+12H^2F_{5X}X^3\ddot{\pi}
\\\nonumber&+6H^2F_{5\pi}X^3 + \,p= 0,
\end{align}
\end{subequations}
{where functions $G_5(\pi,X)$ and $F_5(\pi,X)$ appear upon 
extension of the Lagrangian~\eqref{eq:lagrangian} to the 
most general beyond Horndeski case.}

{The
dynamics of the fluid is governed by the covariant
stress-energy conservation}
\be
\label{eq:SET_cons}
\nabla_{\mu} T^{\mu}_\nu = 0.
\ee

{Turning to perturbations, we
  concentrate on the scalar {modes} 
  (we discuss the tensor
  sector in the subclass of models \eqref{eq:lagrangian} later on;
  this sector is not dangerous in the context of this section).}
  We are going to
derive the linearized field equations
for beyond Horndeski theory with  perfect fluid
and calculate the {propagation speeds for
  two dynamical DOFs present in the scalar sector}.
In what follows we obtain all {expressions}
in a general form first and
then apply them to our model.

{As usual, we make use of} the scalar part of 
ADM parametrization~\eqref{eq:ADM} and the unitary 
gauge approach, so linearized
{gravitational} equations for beyond Horndeski
theory~\eqref{eq:lagrangian} + perfect fluid~\eqref{eq:SET} 
  read\footnote{{The
    scalar component of $(ij)$-equations
  has trace part (which is denoted by $\delta g^{ii}$ here)
  and longitudinal part. The latter, however, is redundant and hence we
  do not use it.}}
\begin{subequations}
\label{eq:linearizedEinstein_rad}
\begin{align}
\label{eq:beta_rad}
& {\delta g^{00}}: \quad \Sigma \alpha - \left(\mathcal{G_T} + \mathcal{D} \dot{\pi}\right) \dfrac{(\triangledown^2\zeta)}{a^2} +3 \Theta \dot{\zeta} - \Theta  \dfrac{(\triangledown^2\beta)}{a^2}
- \frac12\delta \rho  = 0,\\
\label{eq:alpha_rad}
&  {\delta g^{0i}}: \quad \Theta \alpha -\mathcal{{G}_T} \dot{\zeta} - \frac12(p+\rho) \mathcal{V} =0,\\
\label{eq:zeta_rad}
&  {\delta g^{ii}}: \quad  -3 \mathcal{{G}_T} \ddot{\zeta} - 3\left(\dfrac{d \mathcal{{G}_T}}{dt} + 3 H \mathcal{{G}_T}\right) \dot{\zeta} + \mathcal{F_T} \dfrac{(\triangledown^2\zeta)}{a^2} + \left(\mathcal{G_T} + \mathcal{D} \dot{\pi}\right) \dfrac{(\triangledown^2\alpha)}{a^2} + 3 \Theta \dot{\alpha} + 3 (\dot{\Theta} + 3 \Theta H) \alpha \nonumber\\
& 
+\mathcal{{G}_T} \dfrac{(\triangledown^2\dot{\beta})}{a^2} 
+\left(\dfrac{d \mathcal{{G}_T}}{dt} + H \mathcal{{G}_T}\right) \dfrac{(\triangledown^2\beta)}{a^2}
 - \frac32 \,\left(\delta p + (p+\rho)\alpha\right)= 0,
\end{align}
\end{subequations}
where $\delta\rho$, $\delta p$ and $\mathcal{V}$ are perturbations of 
$\rho$, $p$ and the spatial component of
$u_{\mu}$ ($\delta u_{i} = \partial_i \mathcal{V}$), respectively.
 {The coefficients $\Sigma$, ${\cal G_T}$,
  ${\cal D}$ etc. are defined in
  eqs.~\eqref{eq:GS_setup}~--~\eqref{eq:Sigma_coeff_setup} for the
  subclass of models \eqref{eq:lagrangian}, while for general
  beyond Horndeski
theory they are written explicitly in Ref.~\cite{RomaBounce}.}
 {When deriving
  eqs.~\eqref{eq:linearizedEinstein_rad} we used the fact that the
  background
  obeys \eqref{eq:Einstein_rad} and \eqref{eq:SET_cons}.
  Perturbations of
  pressure and energy density are related
in the usual way,
\be
\delta p = u_s^2 \, \delta \rho \; ,
\label{mar31-20-1}
\ee
where $u_s$ is {the} sound speed in the absence of gravity.}
According to the form of eqs.~\eqref{eq:beta_rad} 
and~\eqref{eq:alpha_rad}, $\alpha$ and $\beta$ are 
still non-dynamical {variables.}

Scalar perturbations in the fluid are governed by the linearized
 {covariant conservation
equations~\eqref{eq:SET_cons},}
\begin{subequations}
\label{eq:covariant_rad}
\begin{align}
\label{eq:cov0}
&{\delta\left( \nabla_{\mu}T^{\mu}_ 0\right):}
\qquad
\delta \dot\rho + 3 H (\delta p+\delta\rho) + (p+\rho) 
\left[3\dot\zeta -
\dfrac{\triangledown^2{\beta}}{a^2} - \dfrac{\triangledown^2{\mathcal{V}}}{a^2}\right] = 0, 
\\
\label{eq:covi}
& {\delta\left( \nabla_{\mu}T^{\mu}_ i\right):}
\qquad
\delta p - (p+\rho) \dot{\mathcal{V}} - \left(3H (p+\rho) +\dot{p} + \dot\rho \right) \mathcal{V} +(p+\rho) \alpha = 0.
\end{align}
\end{subequations}

Now we transform the system of eqs.~\eqref{eq:linearizedEinstein_rad}-\eqref{eq:covariant_rad} into the set of two equations, which describe the dynamics of two scalar DOFs,
{namely,} velocity potential $\mathcal{V}$ and 
curvature perturbation $\zeta$. 
First, we make use of constraints~\eqref{eq:beta_rad} 
and~\eqref{eq:alpha_rad} to express non-dynamical $\beta$ and $\alpha$
{through}
$\zeta$, $\delta\rho$ and $\mathcal{V}$. Next, 
{we recall \eqref{mar31-20-1} and 
solve eq.~\eqref{eq:covi}}
for $\delta \rho$. 
Substituting the results into eqs.~\eqref{eq:zeta_rad} 
and~\eqref{eq:cov0},
{we arrive at} the following 
equations 
{for metric perturbation $\zeta$ and
velocity potential $\mathcal{V}$}:
\begin{subequations}
  \label{eq:zeta_v_final}
  \begin{align}
\left( \mathcal{{G}_S} + \dfrac{\mathcal{{G}_T}^2}{\Theta^2}
\dfrac{(p+\rho)}{2\,u^2_s} \right) \ddot{\zeta} -\mathcal{F_S} \dfrac{\triangledown^2{\zeta}}{a^2} - \dfrac{\mathcal{{G}_T}}{\Theta} \dfrac{(p+\rho)}{2\,u_s^2} \ddot{\mathcal{V}} + \dfrac{\left(\mathcal{G_T} + \mathcal{D} \dot{\pi}\right)}{\Theta} \dfrac{(p+\rho)}{2} \dfrac{\triangledown^2{\mathcal{V}}}{a^2} + \dots &= 0,
\\
\dfrac{(p+\rho)}{2\,u_s^2} \left(\ddot{\mathcal{V}} - \dfrac{\mathcal{{G}_T}}{\Theta} \ddot{\zeta} \right) + \dfrac{(p+\rho)}{2} \left( \dfrac{\left(\mathcal{G_T} + \mathcal{D} \dot{\pi}\right)}{\Theta}  \dfrac{\triangledown^2{\zeta}}{a^2} -  \dfrac{\triangledown^2{\mathcal{V}}}{a^2} \right)+ \dots &= 0,
  \end{align}
  \end{subequations}
{where}
we have dropped {terms
  without second derivatives, which}
are irrelevant for the calculation of the propagation speeds 
for $\zeta$ and $\mathcal{V}$.
Let us rewrite
{eqs.~\eqref{eq:zeta_v_final}}
in the matrix form:
\be
\label{eq:matrix_form}
{G_{AB} \,\,\ddot{v}^B} -  F_{AB} \,\,\dfrac{ \nabla_i^2\,{v^B}}{a^2} + \dots = 0,
\ee
where $A,B=1,2$ and $v^1 = \zeta$, $v^2 = \mathcal{V}$. The kinetic matrices
{are} 
\be
G_{AB} = 
\begin{pmatrix}
\mathcal{{G}_S} + \dfrac{\mathcal{{G}_T}^2}{\Theta^2}\dfrac{(p+\rho)}{2\,u^2_s} 
& -\dfrac{\mathcal{{G}_T}}{\Theta}\dfrac{(p+\rho)}{2\,u^2_s} \\
-\dfrac{\mathcal{{G}_T}}{\Theta}\dfrac{(p+\rho)}{2\,u^2_s} & 
\dfrac{(p+\rho)}{2\,u^2_s}
\end{pmatrix},
\;\;
F_{AB} = 
\begin{pmatrix}
\mathcal{{F}_S} 
& -\dfrac{\left(\mathcal{G_T} + \mathcal{D} \dot{\pi}\right)}{\Theta}
\dfrac{(p+\rho)}{2} \\
-\dfrac{\left(\mathcal{G_T} + \mathcal{D} \dot{\pi}\right)}{\Theta}\dfrac{(p+\rho)}{2} & 
\dfrac{(p+\rho)}{2}
\end{pmatrix}.
\ee
{Matrices}
$G_{AB}$ and $F_{AB}$ give the modified stability
conditions for the scalar sector:
{both must be positive-definite, i.e.,
  $\mbox{det}\, G > 0$, $G_{11} >0$,  $\mbox{det}\, F > 0$, $F_{11} >0$.}
  The sound speeds squared are the eigenvalues of the matrix 
$G_{AB}^{-1} F_{AB}$:
  \be
  \label{apr1-20-1}
G_{AB}^{-1} F_{AB} = 
\begin{pmatrix}
\dfrac{\mathcal{F_S}}{\mathcal{{G}_S}} - \dfrac{(p+\rho)}{2\,\mathcal{{G}_S}} \dfrac{\left(\mathcal{G_T} + \mathcal{D} \dot{\pi}\right)\mathcal{{G}_T}}{\Theta^2}  &
-\dfrac{(p+\rho)}{2\,\mathcal{{G}_S}} \dfrac{\mathcal{D}\dot{\pi}}{\Theta}\\
\dfrac{\mathcal{{G}_T}}{\Theta}\left[ 
\dfrac{\mathcal{F_S}}{\mathcal{{G}_S}} - 
\dfrac{(p+\rho)}{2\,\mathcal{{G}_S}} \dfrac{\left(\mathcal{G_T} + \mathcal{D} \dot{\pi}\right)\mathcal{{G}_T}}{\Theta^2}\right] - u_s^2 \dfrac{\left(\mathcal{G_T} + \mathcal{D} \dot{\pi}\right)}{\Theta} \;\;\;&
u_s^2 - \dfrac{(p+\rho)}{2\,\mathcal{{G}_S}} \dfrac{\mathcal{{G}_T}\left(\mathcal{D}\dot{\pi}\right)}{\Theta^2}
\end{pmatrix} \; .
\ee
{They} read as follows:
\bea
\label{eq:speeds_fluid}
c_{\mathcal{S}(1,2)}^2 &=& \dfrac12 u_s^2 +
\dfrac{1}{2} 
\left[ 
\dfrac{\mathcal{F_S}}{\mathcal{{G}_S}} - \dfrac{(p+\rho)}{2\,\mathcal{{G}_S}}\dfrac{\mathcal{{G}_T}(\mathcal{{G}_T}+2\mathcal{D}\dot{\pi})}{\Theta^2} 
\right.\\\nonumber
&&\left.\pm 
\sqrt{ \left(\dfrac{\mathcal{F_S}}{\mathcal{{G}_S}} - \dfrac{(p+\rho)}{2\,\mathcal{{G}_S}}\dfrac{\mathcal{{G}_T}(\mathcal{{G}_T}+2\mathcal{D}\dot{\pi})}{\Theta^2} + u_s^2
\right)^2 - 4 \, u_s^2 \left(\dfrac{\mathcal{F_S}}{\mathcal{{G}_S}} - 
\dfrac{(p+\rho)}{2\,\mathcal{{G}_S}}\dfrac{\left(\mathcal{{G}_T}+\mathcal{D}\dot{\pi}\right)^2}{\Theta^2} \right) } \,\,
\right].
\eea

{Let us pause here to compare these
  formulas with
  known results.
   We recall
  that the coefficient $\mathcal{D}$ vanishes
  in unextended
  Horndeski subclass (see eq.~\eqref{eq:GT_coeff_setup})
  and find that the formulas for kinetic matrices and sound
  speeds given in Ref.~\cite{KobaRev} for Horndeski case
  are restored.
  In particular, for $\mathcal{D} = 0$,
the matrix \eqref{apr1-20-1} is triangular,}
  the matter sound speed 
$c_{\mathcal{S}(2)}^2|_{\mathcal{D}=0}=u_s^2$ is unchanged, while $c_{\mathcal{S}(1)}^2$ of the curvature mode is modified as follows 
(see eq.~\eqref{eq:sound_speeds} for comparison):
\be
\label{eq:speed_Horndeski}
c_{\mathcal{S}(1)}^2|_{\mathcal{D}=0}= \dfrac{\mathcal{F_S}}{\mathcal{{G}_S}} - \dfrac{(p+\rho)}{2\,\mathcal{{G}_S}} \dfrac{\mathcal{{G}_T}^2}{\Theta^2}.
\ee
{On the other hand,}
we note that due to the non-trivial 
coefficient $\mathcal{D}$ in our case, the matrix $G_{AB}^{-1} F_{AB}$
is no longer triangular and both speeds get modified. The latter
{property}
is in agreement with the findings of Refs.~\cite{speed2,speed3}, where similar
observation was made within the EFT approach.

{We now turn to
   our specific Lagrangian 
of
Sec.~\ref{sec:solution},
with
perfect fluid
added}
(see Appendix B for the explicit forms of the Lagrangian functions).
{In the first place, we notice  that
in the class of models \eqref{eq:lagrangian},
  the coefficients $\mathcal{G_T}$, $\mathcal{F_T}$
determining the behavior of tensor perturbations depend on
$\pi$ and $\dot{\pi}$ but do not depend on $H$, see
{eq.~\eqref{eq:GT_coeff_setup}}.
Therefore,
tensor perturbations are not superluminal in the presence of
additional matter if they are not superluminal in pure
beyond Horndeski theory. This is the case in our model of
Sec.~\ref{sec:solution}.}

So, we  analyze
the behaviour of the
{two sound speeds}
~\eqref{eq:speeds_fluid}
{in the scalar sector}.
{We
  consider  {the additional perfect fluid with} the equation of state \[
  p=w\rho \; , \;\;\;\;\; w=u_s^2=\mbox{const}
  \]
for definiteness.}
{We {stress again}
  that our Lagrangian functions are defined only
  in the vicinity
of the line $\dot{\pi}=1$ in the phase space.
Nevertheless,
we keep $\dot{\pi} \neq 1$ at intermediate steps {for generality}.}
Following the analysis strategy adopted
in Sec.~\ref{sec:superluminal}, we 
solve eq.~\eqref{eq:dg00_rad} to
{obtain}
$H_{\pm}(\pi,\dot\pi,\rho)$, where $\pm$
{refers to}
two 
roots of the quadratic equation, cf. the discussion in
the end of Sec.~\ref{sec:solution}.
Then we make use of eq.~\eqref{eq:dgii_rad} and
time derivative of eq.~\eqref{eq:dg00_rad} 
to express $\dot{H}$
and $\ddot{\pi}$ in terms of $\pi$, $\dot{\pi}$, $\rho$ and $\dot\rho$.
Finally, we use the
{covariant
conservation of energy} 
density~\eqref{eq:SET_cons},
$$
\dot{\rho} +3 H (p+\rho) = 0\; ,
$$
 to find $\dot{\rho}$ for given 
$\pi$, $\dot{\pi}$, $\rho$. 
Upon substituting $\dot{H}$, $\ddot{\pi}$ and $H$
into eqs.~\eqref{eq:speeds_fluid}, we get the
{expressions for
  $c_{\mathcal{S}(1,2)}^2(\pi,\dot{\pi},\rho)$. At the last step we
specify to the region in the phase space near $\dot{\pi}=1$.}

\begin{figure}[h!]\begin{center}\hspace{-1cm}
\put(300,50){$\pi$}
{\includegraphics[width=0.6\linewidth] {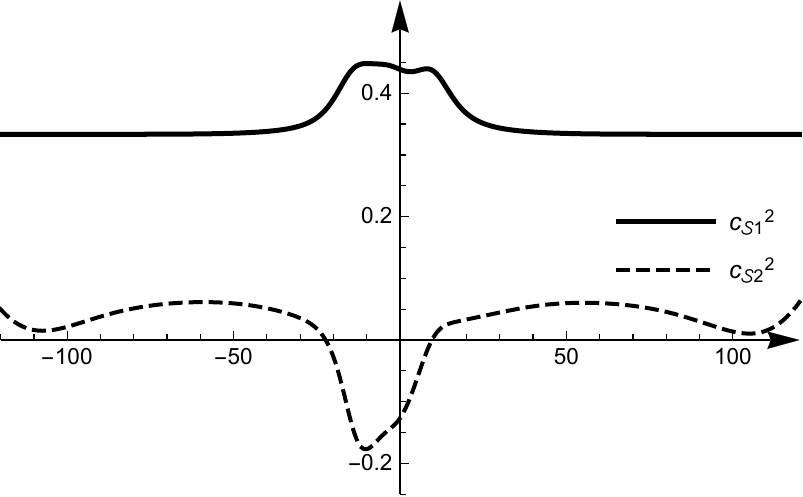}}
\caption{Sound velocities squared at {$\rho=0.1$} (in
    Planck units) as functions of $\pi$ for $w = u_s^2 = 1/3$
    (radiation) at $\dot{\pi}=1$. Negative values of
    $c_{\mathcal{S}(2)}^2$ correspond to unstable region of the phase
    space.}
\label{fig:A}
\end{center}\end{figure}
\begin{figure}[h!]\begin{center}\hspace{-1cm}
\put(300,0){$\rho$}
{\includegraphics[width=0.6\linewidth] {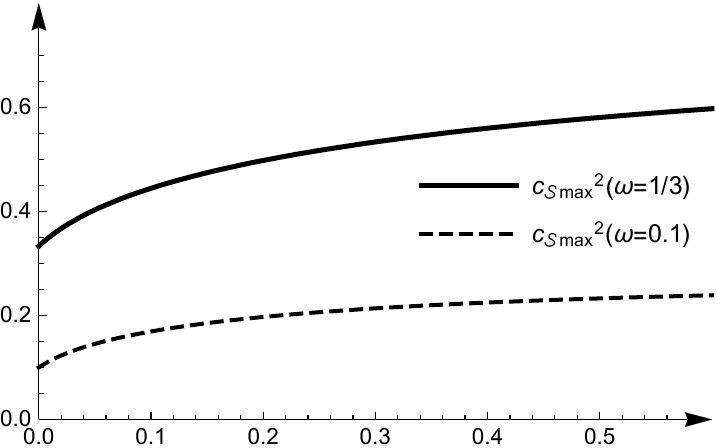}}
\caption{Maximum values of
    the {larger} sound velocity squared,
    $\mbox{max}_\pi \, c_{\mathcal{S}(1)}^2(\pi, \rho)$, as function
of $\rho$
     (in
    Planck units)  for  $w = u_s^2 = 1/3$
    (radiation) and {$w = u_s^2 = 0.1$},
    both at $\dot{\pi}=1$.}
\label{fig:B}
\end{center}\end{figure}

\begin{figure}[h!]\begin{center}\hspace{-1cm}
\put(0,190){$\rho$}
\put(350,0){$\pi$}
{\includegraphics[width=0.7\linewidth] {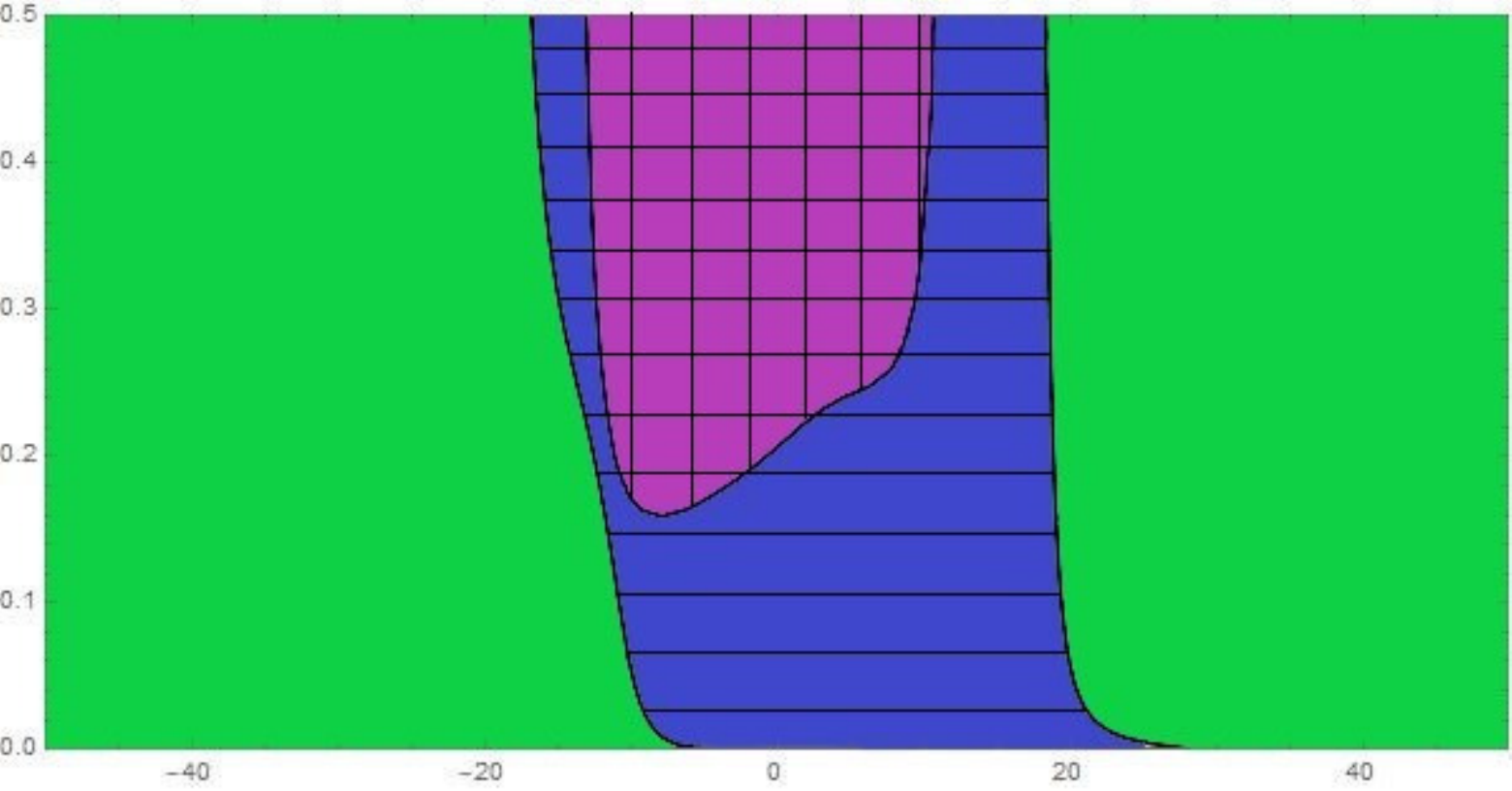}}
\caption{[color online] Phase space $({\pi}, \rho)$ for
  {$w=3/4$},
  where $\rho$ is
  in Planck units. Solid green region is subluminal and stable.
  Instability region ($c_{\mathcal{S}(2)}^2 <0$) is shown in
  blue (horizontal hatching), 
  whereas superluminal region  ($c_{\mathcal{S}(1)}^2 > 1$)
  is shown in purple (vertical hatching). 
  The purple region is inside the blue one,
  so superluminality is actually not problematic.
  }
\label{fig:C}
\end{center}\end{figure}

\begin{figure}[h!]\begin{center}\hspace{-1cm}
\put(0,185){$\rho$}
\put(350,0){$\pi$}
{\includegraphics[width=0.7\linewidth] {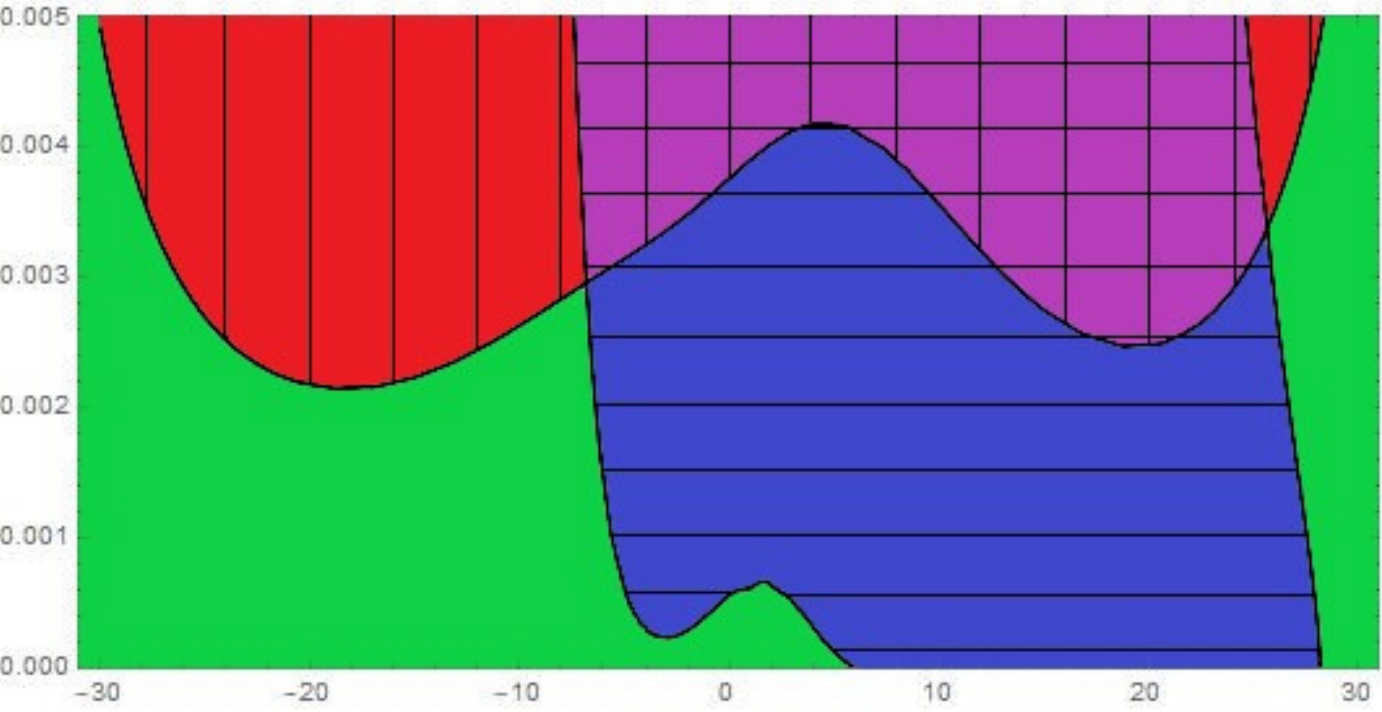}}
\caption{[color online] Same as in Fig.~\ref{fig:C}, but for
  {$w=0.99$}. In the red region (vertical hatching) outside the
  blue one perturbations are stable and superluminal,
$c_{\mathcal{S}(2)}^2 > 0$,
  $c_{\mathcal{S}(1)}^2>1$.  
   }
\label{fig:D}
\end{center}\end{figure}

 {Our results are as follows.
  From the superluminality viewpoint, the dangerous sound speed is
  $c_{\mathcal{S}(1)}$. Even though  $c_{\mathcal{S}(1)}$  may be substantially
      larger than $u_s$ at large $\rho$,  it
    does not exceed 1 in entire phase space
    $(\pi, \rho)$ provided that $w\equiv u_s^2$ is not too large
    (recall that we restrict our analysis to
    the vicinity of $\dot{\pi} = 1$). This is certainly the case
    for $w \leq 1/3$ and somewhat higher.
    We illustrate this situation  in
    Figs.~\ref{fig:A} and
    \ref{fig:B}, which refer to $H_-$  branch. The properties of  $H_+$
    branch are very similar, modulo reflection $\pi \to - \pi$. As $u_s^2$
    increases, there emerge superluminal values of   $c_{\mathcal{S}(1)}$
    at  $\pi$ near zero and large enough $\rho$. At intermediate
    $u_s^2$, however, the superluminality regions are inside
    the instability regions ($c_{\mathcal{S}(2)}^2 <0$), as shown in
    Fig.~\ref{fig:C}. In this case superluminality is harmless.
    As  $u_s^2$ increases even further, the superluminality region
    overlaps partially with the stability region in the phase space.
    This is shown in Fig.~\ref{fig:D}. In this case superluminality is
    genuine, so the values of   $u_s^2$ close to 1 should be avoided
    unless one accepts superluminality. We note that Figs.~\ref{fig:C}
    and \ref{fig:D} show the branch  $H_-$, whereas similar figures,
    with  $\pi \to - \pi$, are obtained for branch $H_+$.
}

{For $u_s^2 = 1$, adding even small amount of matter
   to stable beyond Horndeski cosmology
   makes one of the modes superluminal,
$c_{\mathcal{S}(1)}^2>1$, while keeping the setup
stable, $c_{\mathcal{S}(2)}^2 > 0$. This is the case for any
beyond Horndeski theory and for any cosmological model.
To see this, we write the general expressions for the
the sound speeds \eqref{eq:speeds_fluid} as follows:
\be
c_{\mathcal{S}(1,2)}^2 = \dfrac12 (u_s^2 + \mathcal{A})
\pm  \dfrac12 \sqrt{(u_s^2 - \mathcal{A})^2 + \mathcal{B}}
\nonumber
\ee
where
\be
\mathcal{A}= \frac{\mathcal{F_S}}{\mathcal{G_S}} -
(\rho+p) \, \frac{\mathcal{G_T}(\mathcal{G_T} + 2 \mathcal{D}\dot{\pi})}{2\mathcal{G_S} \Theta^2} \; , \;\;\;\;\;\;
\mathcal{B}= 4u_S^2(\rho+p) \frac{(\mathcal{D} \dot{\pi})^2}{2\mathcal{G_S} \Theta^2} \; .
  \nonumber
  \ee
  Assuming that without matter the setup is stable, one has
$c_{\mathcal{S}(2)}^2 = \mathcal{F_S}/\mathcal{G_S} > 0$
  at $\rho=p=0$, and we consider the case $u_s^2=1$. Adding
  small amount of matter does not ruin the stability
  ($c_{\mathcal{S}(2)}^2 > 0$ by continuity), and,
  since in {\it beyond} Horndeski
  theory one has\footnote{In general beyond Horndeski theory, one
    has~\cite{RomaBounce}
    $\mathcal{D} = 2 F_4(\pi, X) X\dot{\pi} + 6 H F_5(\pi, X) X^2$.
    For given $\pi$ and $\dot{\pi}$, cancellations between the two
    terms are impossible for generic $\rho$, since  the value
    of the Hubble
    parameter depends on $\rho$.}
  $\mathcal{B} >0$, it leads to superluminality, $c_{\mathcal{S}(1)}^2 > 1$
  ($c_{\mathcal{S}(1)}^2 =1 $ for  $\mathcal{B} =0$ and it increases as
  $\mathcal{B}$ increases).  We discuss this point further in
  Sec.~\ref{sec:conclusion}. 
  {Thus,} by continuity, any beyond Horndeski
  cosmological setup becomes superluminal upon adding some matter
  with $u_S^2$ close to 1.  
 }

\section{Conclusion}
\label{sec:conclusion}
We have extended the analysis of the 
bouncing solution~\cite{bounceI} and  studied the phase space 
around the solution. Our analysis has shown that, although 
the solution itself is
free of pathological DOFs and superluminal modes 
during entire evolution, there is a
region with superluminal gravitational 
waves in its close neighborhood. Thus, the original model~\cite{bounceI}
suffers the
superluminality problem.

To cure this,
we have suggested a new version of the 
bouncing scenario, where the propagation speed of the tensor modes is 
strictly smaller than the speed of light during and around the 
bouncing epoch.
As a result, there are no superluminal regions {in the} phase space
close to the new solution.
By analysing the behaviour of the bouncing trajectories with different
initial conditions we have also checked
that our construction does not involve
fine-tuning.
{In the early- and late-time
  asymptotics, $t\to - \infty$ and $t\to +\infty$,
  the model reduces to General
  Relativity and conventional massless scalar
  field driving contraction and expansion, respectively.}

{We have then considered a more
  general issue of possible superluminality in 
  {our beyond Horndeski} model with
  extra ideal fluid {added}. We have seen that as long as the flat-space
  sound speed
  in additional matter is not close to 1, all perturbations remain subluminal
  in the region of {the} phase space (near $\dot{\pi}=1$) where
  the Lagrangian of our model is explicitly known. This is the case,
  in particular, for additional matter in the form of radiation.}

{On the other hand, we have shown
  that our model and, {in fact}, any beyond
  Horndeski model (in {a} cosmological setting) 
  becomes superluminal\footnote{We have shown this explicitly for
    beyond Horndeski theory with one scalar field (Galileon),
    but it is likely that this property
    holds for theories with several Galileons.}
  upon adding even small energy density of
  extra matter with luminal flat-space
  sound speed, $u_s= 1$ (or almost luminal $u_s$).
  As we pointed out in Introduction,
  once one insists on the absence of superluminality,
   the latter observation has interesting consequences for
   scalar-tensor theories  with several scalar fields, at least one
   of which is 
   in  beyond Horndeski regime. Namely, neither of
   these scalars can have canonical kinetic term and minimal coupling to
   metric, {since} otherwise the flat-space sound speed
   of a scalar field would be equal to 1, that field could
   have small but non-zero
   homogeneous energy density and, therefore, at least one of the excitations
   about FLRW background would be superluminal.}

Coming back to our novel model, we conclude that it may
be considered as another step towards 
constructing application-oriented cosmologies. 
Complete stability, GR asymptotics and subluminality
of the new bouncing solution make
it a promising candidate for future realistic models of 
the early Universe.

\section*{Acknowledgements}
The authors are grateful to S. Dubovsky and 
R. Rattazzi for helpful discussions
and to A. Vikman, G. Ye and A. Anabalon for useful correspondence.
          {The authors thank the anonymous referee for}
           {valuable comments and for}
            {pointing out 
Ref.~\cite{BKLO} in the context of $\gamma$-crossing phenomenon.}
The work of S.M. on Sec.~2 has been supported by the Foundation 
for the Advancement of Theoretical Physics and Mathematics 
“BASIS” grant, the work on
{Secs.~3,~4} has been supported by 
Russian Science Foundation grant 19-12-00393.

\section*{Appendix A}
\label{sec:AppendixA}
In this Appendix we give an example set of Lagrangian 
functions~\eqref{eq:lagr_series} for a healthy bouncing 
model suggested in Ref.~\cite{bounceI}: 
\be
\label{g40_g41_old}
g_{40}(\pi) = - g_{41}(\pi) = - \frac{w}{2} \;\mbox{sech}^2\left(\frac{\pi}{\tau}+u\right),
\ee
\be
\label{f_40_old}
f_{40}(\pi) = -f_{41}(\pi) + w\; \mbox{sech}^2\left(\frac{\pi}{\tau}+u\right),
\ee
\be
\label{f_41_old}
f_{41}(\pi) = \frac{3w\; \mbox{sech}^2\left(\frac{\pi}{\tau}+u\right)}{2\pi\tau}\; \left[ \pi^2\; \mbox{tanh}\left(\frac{\pi}{\tau} + u\right) +\tau^2\;\mbox{tanh}\left(\frac{\pi}{\tau}\right)-\pi\;\tau \right],
\ee
\bea
& f_0(\pi)& = \frac{1}{3 \tau \left(\pi^2+\tau^2\right)^2}\; \Bigg[-\tau^3+3 \tau \left(\pi^2+\tau^2\right)^2\; f_2(\pi)-3 \pi \tau^2 w \;\mbox{sech}^2\left(\frac{\pi}{\tau}+u\right) \mbox{tanh}\left(\frac{\pi}{\tau}\right)\\\nonumber
&&+3 \pi^3 w\; \mbox{sech}^2\left(\frac{\pi}{\tau}+u\right) \mbox{tanh}\left(\frac{\pi}{\tau}+u\right)+6 \pi \tau^2 w\; \mbox{sech}^2\left(\frac{\pi}{\tau}+u\right) \mbox{tanh}\left(\frac{\pi}{\tau}+u\right)\Bigg],
\\
\label{explicit_f1_old}
& f_1(\pi) &= 
-\frac{1}{3 \tau \left(\pi^2+\tau^2\right)^2}\; \Bigg[\tau^3+6 \tau (\pi^2+\tau^2)^2 \; f_2(\pi)-3 \pi \tau^2 w \; \mbox{sech}^2\left(\frac{\pi}{\tau}+u\right) \mbox{tanh}\left(\frac{\pi}{\tau}\right)\\\nonumber
&&+3 \pi^3 w\; \mbox{sech}^2\left(\frac{\pi}{\tau}+u\right) \mbox{tanh}\left(\frac{\pi}{\tau}+u\right)+6 \pi \tau^2 w\; \mbox{sech}^2\left(\frac{\pi}{\tau}+u\right) \mbox{tanh}\left(\frac{\pi}{\tau}+u\right) -\pi^2 \tau \Bigg],
\\
&f_2(\pi) & =\frac{1}{12 \tau (\pi^2+\tau^2)^2}\;\left\lbrace\tau^3+3 \tau^5+4 \pi w\; \mbox{sech}^2\left(\frac{\pi}{\tau}+u\right) \left[-4 \pi \tau +9 \tau^2 \;\mbox{tanh}\left(\frac{\pi}{\tau}\right)\right.\right. \\\nonumber
&&\left.\left.+3 (\pi^2-2 \tau^2) \;\mbox{tanh}\left(\frac{\pi}{\tau}+u\right)\right]\right\rbrace,
\eea
where $u$ and $w$ are numerical parameters. The bouncing solution is given by eq.~\eqref{eq:Hubble} and has $\dot{\pi}=1$.
The plots in Sec.~\ref{sec:superluminal} are given for $u=1/10$,
$\tau=10$, $w=1$.

\section*{Appendix B}
\label{sec:Appendix}
Here we give an explicit example  of the Lagrangian
functions~\eqref{eq:lagr_series-bis} defining
the beyond Horndeski theory~\eqref{eq:lagrangian},
which admits a new completely stable bouncing solution
with no superluminal modes in the vicinity of the solution:
\be
g_{40}(\pi) = - g_{41}(\pi)  = -\dfrac12\; \mbox{sech}^2\left(\frac{\pi}{\tau}+u\right),
\ee
\be
f_{40}(\pi)=-f_{41}(\pi)+\dfrac{5w}{4}\;\mbox{sech}\left(\frac{\pi}{\tau}+u\right),
\ee
\bea
&f_{41}(\pi)& = \dfrac{1}{(8 \pi \tau)} \;
\mbox{sech}\left(\frac{\pi}{\tau}+u\right) 
\Bigg[-25 \pi \tau w + 8 \pi \tau \; \mbox{sech}\left(\frac{\pi}{\tau}+u\right)\\&&\nonumber
+12 \pi^2 \; \mbox{sech}\left(\frac{\pi}{\tau}+u\right)  
\mbox{tanh}\left(\frac{\pi}{\tau}+u\right) +12 \tau^2 \mbox{sech}\left(\frac{\pi}{\tau}+u\right)  \mbox{tanh}\left(\frac{\pi}{\tau}\right) \Bigg],
\eea
\bea
&f_1(\pi)&
 = \dfrac{1}{12 \tau (\pi^2+\tau^2)^2} \Bigg[
4 \tau (\pi^2-3 \tau^2)+12 \tau (\pi^2+\tau^2)^2\; f_3(\pi)
\\&&\nonumber
+15 w \;\mbox{sech}\left(\frac{\pi}{\tau}+u\right) \Big(\tau (\pi^2-2 \tau^2)+2 \pi (\pi^2+\tau^2)\; \mbox{tanh}\left(\frac{\pi}{\tau}+u\right)\Big)
\\&&\nonumber
-12\;\mbox{sech}^2\left(\frac{\pi}{\tau}+u\right) \Big(\tau (\pi^2-2 \tau^2)+\pi \tau^2 \; \mbox{tanh}\left(\frac{\pi}{\tau}\right)+ \pi (3 \pi^2+2 \tau^2) \; \mbox{tanh}\left(\frac{\pi}{\tau}+u\right)\Big)
\Bigg],
\\
&f_2(\pi)&
 = \dfrac{1}{12 \tau (\pi^2+\tau^2)^2} \Bigg[
4 \tau^3-24 \tau (\pi^2+\tau^2)^2 \;f_3(\pi)
\\&&\nonumber
- 5 w \;\mbox{sech}\left(\frac{\pi}{\tau}+u\right)
\Big(\tau (\pi^2-2 \tau^2)+2 \pi (\pi^2+\tau^2)
\;\mbox{tanh}\left(\frac{\pi}{\tau}+u\right) \Big)
\\&&\nonumber
+ 4\; \mbox{sech}^2\left(\frac{\pi}{\tau}+u\right)
\Big[ 3 \pi \tau^2 \;\mbox{tanh}\left(\frac{\pi}{\tau}\right)+(\pi^2-2 \tau^2) \left(\tau+\pi \;\mbox{tanh}\left(\frac{\pi}{\tau}+u\right)\right)\Big]
\Bigg],
\\
&f_3(\pi)&= \dfrac{1}{24 \tau (\pi^2+\tau^2)^2} \Bigg[
6 \tau^3 (-1+\tau^2)
\\&&\nonumber
+5 w \;\mbox{sech}\left(\frac{\pi}{\tau}+u\right)
\Big(-16 \pi^2 \tau-3 \tau^3+3 \pi (\pi^2+\tau^2)
\;\mbox{tanh}\left(\frac{\pi}{\tau}+u\right)\Big)
\\&&\nonumber
+4 \; \mbox{sech}^2\left(\frac{\pi}{\tau}+u\right)
\Big[8 \pi^2 \tau+3 \tau^3+12 \pi \tau^2 \;\mbox{tanh}\left(\frac{\pi}{\tau}\right)+6 \pi (\pi^2-\tau^2)
\;\mbox{tanh}\left(\frac{\pi}{\tau}+u\right)\Big]
\Bigg].
\eea
The bouncing solution is given by eq.~\eqref{eq:Hubble} and has 
$\dot{\pi}=1$.
The plots in Sec.~\ref{sec:solution} and Sec.~\ref{sec:extra_matter} are given for $u=1/10$,
$\tau=10$, $w=1$.

\end{document}